\documentclass[11pt,a4paper,usenatbib]{emulateapj}
\usepackage{color}
\pdfoutput=1

\newcommand{\BT}{\rm\,B_T} 
\newcommand{\ugriz}{\,ugriz}
\newcommand{\Msol}{\rm\,M_{\odot}} 
\newcommand{\Mstellar}{\rm\,M_{*}}
\newcommand{\Mhalo}{\rm\,M_{halo}} 

\newcommand{\Msolyr}{\rm\,M_{\odot}\,yr^{-1}}
 
\newcommand{\kpc}{\rm\,kpc}

\newcommand{\kms}{\rm\,km\,s^{-1}}
\newcommand{\MB}{\ensuremath{M_{B}}}

\newcommand{\Plog}{\ensuremath{P_{\beta = 0}}}

\shorttitle{RC3-SDSS Galaxy Morphology and Environment}
\shortauthors{Wilman and Erwin}

\begin{document}

\title{The Relation Between Galaxy Morphology and Environment in the
Local Universe: An RC3-SDSS Picture}

\author{David~J.~Wilman\altaffilmark{1}, Peter Erwin\altaffilmark{1,2}}

\altaffiltext{1}{Max-Planck-Insitut f\"{u}r extraterrestrische Physik,
Giessenbachstrasse, 85748 Garching, Germany}
\altaffiltext{2}{Universit\"{a}ts-Sternwarte M\"{u}nchen,
Scheinerstrasse 1, 81679 M\"{u}nchen, Germany}

\begin{abstract}

We present results of an analysis of the local ($z \sim 0$)
morphology-environment relation for 911 bright ($M_B < - 19$) galaxies, based on
matching classical RC3 morphologies with the SDSS-based group catalog of Yang et
al., which includes halo mass estimates. This allows us to study how the
relative fractions of spirals, lenticulars, and ellipticals depend on halo mass
over a range of $10^{11.7}$--$10^{14.8} \, h^{-1} \Msol$, from isolated
single-galaxy halos to massive groups and low-mass clusters. We pay particular
attention to how morphology relates to central vs.\ satellite status (where
``central'' galaxies are the most massive within their halo). The fraction of
galaxies which are elliptical is a strong function of stellar mass; it is also a
strong function of halo mass, but \textit{only} for central galaxies. We
interpret this as evidence for a scenario where elliptical galaxies are always
formed, probably via mergers, as central galaxies within their halos, with
satellite ellipticals being previously central galaxies accreted onto a larger
halo. The overall fraction of galaxies which are S0 increases strongly with halo
mass, from $\sim 10$\% to $\sim 70$\%.  Here, too, we find striking differences
between the central and satellite populations. $20 \pm 2$\% of central galaxies
with stellar masses $\Mstellar > 10^{10.5}\Msol$ are S0 regardless of halo mass,
but satellite S0 galaxies are only found in massive ($> 10^{13} \, h^{-1}
\Msol$) halos, where they are $69 \pm 4$\% of the $\Mstellar >
10^{10.5}\Msol$ satellite population.  This suggests \textit{two} channels for
forming S0 galaxies: one which operates for central galaxies, and another which
transforms lower mass ($\Mstellar \lesssim 10^{11}\Msol$) accreted spirals into
satellite S0 galaxies in massive halos.  Analysis of finer morphological
structure (bars and rings in disk galaxies) shows some trends with stellar mass,
but none with halo mass; this is consistent with other recent studies which
indicate that bars are not strongly influenced by galaxy environment.  Radio
sources in high-mass central galaxies are common, similarly so for elliptical
and S0 galaxies, with a frequency that increases with the halo mass.
Emission-line AGN (mostly LINERs) are more common in S0s, but show no strong
trends with environment.

\end{abstract}


\section{Introduction}\label{sec:intro}

Galaxies come in many different shapes and sizes, but primarily consist of two
dynamically stable components, bulges and disks, with additional contributions
from quasi-stable features such as bars and rings. \citet{Hubble26} devised what
has become known as the ``tuning fork'' classification to describe this
morphological schema: elliptical galaxies ($\sim$ pure bulge); spiral galaxies
(disks containing spiral features, both barred and unbarred, with a sequence
Sa--Sc\footnote{\citet{RC1} extended this sequence further to Sd and Sm} of
decreasing bulge component and increasing spiral arm opening angle); and
lenticular ``S0'' galaxies (defined by the presence of a disk with no
discernable spiral arms).

The abundance of these galaxy types is now known to correlate strongly with
environment: elliptical galaxies live preferentially in regions of very high
local density \citep{Dressler80}, inhabiting the cores of clusters and groups
rather than their outskirts \citep{Melnick77,Whitmore91,Wilman09}. However, the
total luminosity-limited fraction of ellipticals is similar in a wide range of
environments \citep{Desai07,Wilman09,Just10} and has evolved very weakly since
$z \sim 1$ \citep{Dressler97,Fasano00,Smith05,Postman05}.

Lenticular galaxies, on the other hand, are globally no more abundant in
clusters than they are in groups \citep{Wilman09}, with a much weaker dependence
on local density than ellipticals
\citep{Dressler80,Dressler97,Postman05,Poggianti08}. A lower fraction at fixed
luminosity is found only in the lower density field \citep{Wilman09}.  The
global fraction of lenticular galaxies has grown by a factor $\gtrsim 2.5$ in
groups and clusters since $z \sim 0.5$ \citep{Dressler97,Fasano00,Wilman09},
potentially more rapidly in groups and low mass clusters than in high-mass
clusters \citep{Poggianti09,Just10}.  Since lenticulars and ellipticals make up
the bulk of the low redshift passive population \citep[e.g.][]{Bundy10}, this
means that the majority of galaxies which have ceased forming stars since $z
\sim 0.5$ have retained their disks as lenticulars.  Beyond $z \sim 0.5$ (up to
$z \sim 1$), there is no evidence for further evolution in the elliptical or S0
fraction in the high-mass clusters sampled to date \citep{Smith05,Postman05}.

To build a global picture of galaxy evolution, it is important to
understand how the evolution of galaxies and their environment is
intertwined, and which physical processes drive the suppression of star
formation and the morphological transformations which are required to
explain these observations.

It is attractive to explain morphological evolution in the context of bulge
growth through galaxy mergers \citep{Springel05,Hopkins10}, which take place in
the center of halos as a natural consequence of hierarchical clustering and
dynamical friction \citep[e.g.][]{deLucia07,deLucia11}.  Elliptical galaxies
almost certainly form through a mixture of major and minor mergers.  However,
the role of mergers in the formation of bulges in galaxies with disks is less
clear.  Cooling flows should form at the center of halos, and will reform a disk
around the merger remnant so long as the gas is not too efficiently reheated. 
SPH simulations suggest that during mergers of galaxies with high gas fractions,
much of the gas retains its angular momentum and reforms a disk, and that some
component of the stellar disk can survive in all but 1:1 mergers
\citep{Hopkins09}. Meanwhile, secular processes might contribute $\lesssim 10\%$
of the galaxy mass in a ``pseudobulge'' \citep{Kormendy04}, or potentially a
larger fraction at high redshift where disks are less dynamically stable
\citep{Genzel08}.  That S0 galaxies exhibit more significant bulges on average
than spiral galaxies \citep[e.g.][]{Dressler80,Wilman09,Laurikainen10} is
consistent with a merger origin.  Nonetheless, there is currently no more than
circumstantial evidence that bulge growth and the suppression of star formation
in most S0s are causally linked.

Our observational picture still misses one vital and surprisingly
straightforward ingredient: a full picture of how galaxy morphology depends upon
environment at low redshift.  Early studies of environment focussed almost
exclusively on galaxy clusters, for which good statistics and strong trends are
to be found even without high spectroscopic completeness.  The more recent
emphasis has been on extending these trends to high redshift with high
resolution Hubble Space Telescope (HST) imaging.  Modern local surveys such as
the Sloan Digital Sky Survey (SDSS) are so large that much effort has been
invested in the automated classification of morphological properties, such as
the simple SDSS ``concentration'' index, and multi-component decomposition with
different levels of sophistication \citep[resolution dependent,
e.g.][]{Allen06,Gadotti09}. An interesting alternative solution is the ``Galaxy
Zoo'', providing visual classifications for the whole SDSS, by employing an
enthusiastic public to classify galaxies \citep{Lintott11}.

The disadvantage of all these approaches is that new classification schemes tend
to be relatively simplistic, with unknown systematics, and produce data which is
difficult to assess on the basis of our existing understanding of the Hubble
scheme.  Standard Hubble type classifications within SDSS are available only for
subsamples \citep[e.g.][]{Fukugita07}, although these are becoming larger
\citep{Nair10}.

In this paper we have taken the simple step of taking a sample of
galaxies from the Third Reference Catalog of Bright Galaxies
\citep[RC3,][]{RC3} which is matched to the SDSS Data Release 4 (DR4)
to provide classical morphological information for a large sample of
galaxies for which the selection function and environment can be
properly characterized.  For our purposes this provides an ideal sample
to study environmental trends for galaxies with detailed and well
understood classifications.

In section~\ref{sec:sample} we describe the sample, discussing the selection
function of RC3 galaxies and the group catalog which we use to describe galaxy
environment.  Section~\ref{sec:results} presents our results, including Hubble
type fraction as a function of stellar and halo mass, a dichotomy between
central and satellite galaxies, the nature of activity in the SDSS fiber spectra
of galaxies with different morphological types and environments, and the
dependence (or lack thereof) of morphological components (bars, rings etc) on
environment. This is further discussed in the context of a hierarchically
evolving Universe in section~\ref{sec:disc}, and in section~\ref{sec:concl} we
present our conclusions.

To compute distances, absolute magnitudes, and stellar masses, we assume a
$\Lambda$CDM cosmology with $\Omega_M$, $\Lambda$, and $H0$ equal to $0.3$,
$0.7$, and 71 km~s$^{-1}$ Mpc$^{-1}$, respectively. Halo masses are presented in
$h^{-1} \Msol$, to retain the native units of \citet{Yang07}.\footnote{$h=0.71$
in our adopted cosmology.}

\section{Sample}\label{sec:sample}

\subsection{The RC3 Catalog}\label{sec:rc3}

The RC3 catalog (Third Reference Catalog of Bright
Galaxies)\footnote{http://heasarc.nasa.gov/W3Browse/all/rc3.html}
provides information for a large sample of nearby galaxies \citep{RC3}.
It ``attempts to be reasonably complete for galaxies having apparent
diameters larger than 1 arcmin at the D25 isophotal level and total {\it
B}-band magnitudes $\BT$ brighter than about 15.5, with a redshift not
in excess of 15,000 km/s''. Some smaller, fainter, and more distant
galaxies are also included. The most important aspect from our point of
view is that detailed morphological classifications (Hubble types) from
photographic plates are provided.

Inspection of SDSS imaging for a subsample of RC3 galaxies confirms the
majority of classifications.  However, a significant minority of
galaxies appear to be falsely classified, usually in the sense that some
galaxies appearing to be early-type spirals or ellipticals are instead
classified S0.  We describe our reclassification process in
section~\ref{sec:reclass}.

\subsection{The SDSS Catalog}\label{sec:sdss}

The SDSS DR4 \citep[Sloan Digital Sky Survey Data Release 4,
][]{SDSSDR4} provides $\ugriz$ photometry and spectroscopy for 565,715
galaxies across a total of 4783 square degrees. In the {\it main}
sample, this is highly complete down to a limiting magnitude of 17.77 in
$r$-band dereddened Petrosian magnitudes, and limiting surface
brightness of $\mu_r = 23.0$ mag arcsec$^{-2}$. Low level incompleteness
exists mainly in the highest density regions due to the inability to
assign fibers to targets with separations $ < 55\arcsec$.

The SDSS spectroscopic sample is incomplete at the bright end due to fiber
magnitude limits, applied to avoid saturation and excessive cross-talk in the
spectrographs.  Our catalog is based on the DR4 version of the New York
University Value Added Galaxy Catalog \citep[NYU-VAGC][]{Blanton05}), which
includes objects with bright fiber magnitudes and is photometrically
recalibrated across the sky. If we restrict ourselves to the SDSS area with DR4
spectroscopic coverage, we measure redshift completenesses of $\sim 93$\% for $r
< 16$, $\sim 88$\% for $13 < r < 14$, and $\sim 80$\% for $12 < r < 13$ (i.e.,
the fraction of galaxies in the NYU-VAGC which actually have SDSS spectroscopy).
\citet{Strauss02} studied the incompleteness of SDSS galaxy spectroscopy using
local catalogs and found typical incompleteness of $\sim 5$\% for bright ($r
\lesssim 15$) galaxies. Significantly, this incompleteness was due to overlaps
with saturated stars. What this indicates is that the bright-galaxy
incompleteness is both low and, crucially, not related to galaxy type. Therefore
this incompleteness is not important for our analysis.

\subsection{A Matched RC3-SDSS Catalog}\label{sec:matchcat}

We use the NYU-VAGC match to RC3 galaxies from \citet{Blanton05}.
Although the matching radius was 45\arcsec\ to compensate for the
variable RC3 astrometrical accuracy, all matches are within 15\arcsec\
\citep[figure~7 of][]{Blanton05}.  Most of the galaxies under
consideration are themselves larger than 15\arcsec\ in size, and thus
the incidence of false matches is low. There may be few occasions in
which a false match is made for interacting or neighbouring galaxies,
but this is unlikely to greatly influence our overall statistics.

Our full matched RC3-SDSS catalog contains 1340 galaxies, mostly in the
range $0.01 < z < 0.04$ (median $z = 0.023$), with a tail of bright
galaxies extending to $z = 0.13$.

\subsection{Morphological Reclassification}\label{sec:reclass}

Subsequent examination of individual galaxies suggested that some of the RC3
classifications were in error (e.g., a ``spiral'' galaxy that was clearly an
elliptical, or an ``S0'' that had strong spiral arms). To address this, we
carried out a reclassification analysis for the sample. Two student interns
independently examined all galaxies in our sample with assigned halo masses,
identifying 406 cases with potentially incorrect general classification (i.e. E,
S0 or spiral). We then examined each of these objects in detail, reclassifying
them if necessary.\footnote{For reclassification, both color JPEG and
background-subtracted $g$-band FITS images were examined.} In cases where a
lenticular was reclassified as a spiral or vice-versa, we carried over any
original disk-structure notations (bars and rings) from the original RC3
classification.  165 of the 406 flagged galaxies ended up with new
classifications, including 6 of 23 ellipticals, 80 of 129 S0s and 65 of 230
spirals (a few galaxies have irregular or merger types). One galaxy (UGC5677)
was clearly matched to the wrong SDSS object, due to especially poor RC3
astrometry; we assigned it zero weight in order to remove it from the sample. 
Finally, we also classified a total of 55 galaxies with $B \leq 16$ (our primary
magnitude limit; see Section~\ref{sec:selection}) which had \textit{no}
classifications in the RC3.

The most striking aspect of this analysis was the high fraction of
reclassified lenticulars: a total of 40\% of the original (RC3)
lenticulars ended up with different Hubble types (16 as ellipticals, 34
as spirals, and 9 as peculiar/merger-remnant systems). A similar high
proportion of reclassified RC3 lenticulars can be seen in Figure~1 of
\citet{Fukugita07} and Figure~14 of \citet{Nair10}. This led to a
general reduction in the overall fraction of S0s and smaller increases
in the elliptical and spiral fractions. The reclassification process
actually strengthened the trends described in section~\ref{sec:results}.

\subsection{SDSS Photometry and the RC3 Selection Function}\label{sec:selection}

Our goal is to examine the morphological composition of galaxies using
the corrected RC3 classifications.  This requires a detailed
understanding of the inherently non-uniform RC3 selection, using the
full SDSS catalog as the nominally complete ``parent sample''.

Our photometry is based on the {\it model} magnitudes in the NYU-VAGC
catalog, which we have corrected for galactic extinction
\citep{Schlegel98}. For large galaxies ($r_{50} \gtrsim 10\arcsec$,
where $r_{50}$ is the radius containing half the $r$-band Petrosian
flux), the SDSS background can be {\it oversubtracted}, due to the outer
part of the galaxy being treated as part of the sky; this effect is
approximately color independent \citep{Blanton11}. To account for this,
we applied the photometric correction published by Blanton et al., as a
function of the measured half-light radius $r_{50}$ (their Table~1,
correction to {\it v5.4} sky). This correction is generally consistent
with those published by \citet{West10} and \citet{HydeBernardi09}. A
comparison with the latter correction for ellipticals -- Figure~14 of
\citet{Blanton11} -- suggests that these corrections \textit{might}
depend upon morphology.  However, the differences are only significant
($d(mag) \sim 0.5$) for galaxies with $r_{50} \gtrsim 25\arcsec$. Since
only one of the galaxies in our sample is this large, we do not consider
morphology-dependent corrections to be something to worry about.

As noted above, the main selection functions applying to the RC3 catalog
are limits on (photographic) blue magnitude and isophotal diameter. To
determine how the magnitude limit affected the selection, we generated
an equivalent $B$ magnitude for all SDSS galaxies, using the SDSS $u$
and $g$ magnitudes\footnote{http://www.sdss.org/dr7/algorithms/sdssUBVRITransform.html \#Lupton2005}: 
\begin{equation} 
B = u - 0.8116(u - g) + 0.1313. 
\end{equation} 
To evaluate the effects of the diameter
limit, we used the SDSS $r_{90}$ measurement (the radius containing 90\% of the
Petrosian flux in r-band).

How well do our SDSS-derived $B$ magnitudes match the (largely
photographic) magnitudes used as limits for RC3? Measurements of the RC3
total $B$-band magnitude, $B_{T}$, are available for a sub-sample of 150
SDSS-RC3 galaxies. For these objects we compare $B_{T}$ to $B$ derived
from the SDSS photometry. We find that the RC3 and SDSS $b$ magnitudes
are comparable in 144/150 cases, with significant scatter due to the
large RC3 photometric errors.  The other 6 objects have significantly
underestimated fluxes from SDSS, with measured $g$-band magnitudes of $g
\geq 15.8$ (corresponding to $B \gtrsim 16.0$). Five of these are
clearly cases where deblending-related problems have affected the SDSS
measurements.  The sixth is an RC3 object (NGC842) mis-matched to the
wrong SDSS galaxy due to a $\sim 1\arcmin$ error in its RC3 cataloged
position. We remove these objects by only considering galaxies with $g
\leq 15.8$ and $B \leq 16.0$, although the ``true'' magnitudes might be
brighter than this limit. Assuming that this analysis is representative,
we estimate that $\sim 4$\% of objects are lost due to poor SDSS
photometry, and a $< 1$\% occurance of RC3-SDSS mismatches.

\begin{figure*}
  \epsscale{1.00}
  \plotone{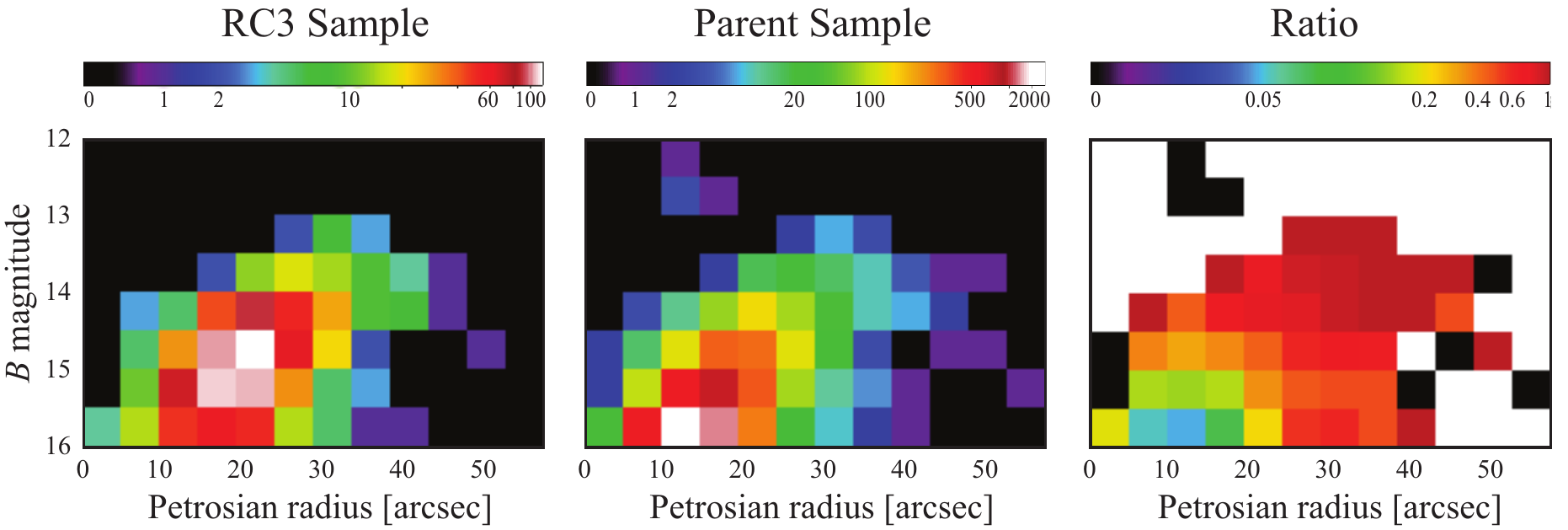}
  \caption{2D histograms used to determine the RC3 selection function,
    using a $B = 16$ limit. The left and middle panels show counts of
    RC3 galaxies and SDSS parent sample galaxies binned by $r_{90}$
    (the radius containing 90\% of the $r$-band Petrosian flux) and
    $B$-band magnitude; horizontal color bars indicate the scales in
    galaxies per bin. The right panel shows the ratio of the left and
    middle panels, which is the basis of the RC3 selection
    function. The horizontal color bar indicates the ratio (from 0.0
    to 1.0); white bins indicate no data (i.e., no galaxies in either
    sample are found in those bins).}
  \label{figure:rc3selection}
\end{figure*}

The first two panels of Figure~\ref{figure:rc3selection} illustrates how
SDSS-RC3 galaxies and the (almost complete) SDSS parent sample populate
the magnitude-size parameter space. Not surprisingly, fainter or smaller
galaxies are less likely to be found in the RC3 catalog. The third panel
shows the ratio by number of RC3 to parent-sample objects as a function
of these two parameters. To correct for the RC3 selection biases, we
weight SDSS-RC3 galaxies by the inverse of this selection function:
$weight(B,r_{90}) = \frac{N_{parent}(B,r_{90})}{N_{RC3}(B,r_{90})}$,
computed in bins of 0.5 mag in $B$ magnitude and 5\arcsec\ in $r_{90}$.
Some of the fainter, smaller objects have weights of up to 41.
A threshold brighter than $B = 16.0$ would remove these objects, but
significantly worsen the overall statistics. We examine the robustness
of our results by applying three different limits -- $B \leq 16.0$,
15.5, and 15.0 -- and checking that each result is consistent with each
of the three limits applied.

We would like to compute fractions of the complete population down to a
given luminosity limit. However, more luminous galaxies are visible to
larger distances, and thus over larger cosmological volumes (this is
the well-known Malmquist bias). To correct for this, we compute
$V/V_{\rm max}$ weights for our galaxies. In practice, $V_{\rm max}$ is the
volume within which a given galaxy is visible, and $V$ is the volume of
the survey, defined to be the maximum value of $V_{\rm max}$ (minimum
$V/V_{\rm max} = 1$). To ensure we are insensitive to large $V/V_{\rm max}$
weights, we have carefully examined our results both with and without
these weights applied, and as a function of luminosity. Robust results
are achieved with a luminosity limit of $\MB = -19.0$.

\subsection{The Group Catalog}\label{sec:groups}

For a full picture of how galaxy morphologies depend upon environment,
we use the group catalog of \citet[][Y07]{Yang07} constructed from 
SDSS DR4, and applied to galaxies in the range $0.01 \leq z \leq 0.2$.
The Y07 catalog is constructed iteratively using an algorithm which is
an amalgam of friends-of-friends linking of galaxies to form groups and
abundance matching to assign masses to those groups.  Group halo masses
are assigned based upon the rank order in terms of the group total
stellar mass or luminosity of all galaxies brighter than an evolution
and k-corrected $r$-band absolute magnitude of $-19.5$.  We use the
stellar-mass-based halo masses, for which halos are complete down to
$\Mhalo = 10^{11.63} \, h^{-1} \Msol$; 
this corresponds to a single red galaxy of stellar mass $\Mstellar =
10^{10} \, \Msol$ just making the $-19.5$ cut in luminosity. This method
provides halo masses for single as well as grouped galaxies.

We use {\it Sample II} as described by Y07, which contains a total of
369,447 galaxies including those with redshifts from sources other than
SDSS and assumed redshifts for fiber collision galaxies as described.
Galaxies and groups with incomplete local information are removed from
the catalog, which means galaxies with completeness in the local
``tiling sector'' \citep{Blanton05} of C~$ < 0.7$ (missing $\geq 30\%$
of neighbours) and groups on the edge of the survey with completeness
$f_{edge} < 0.6$ ($\geq 40\%$ of group members with C~$ < 0.7$).
Isolated galaxies with stellar masses $\Mstellar < 10^{10} \Msol$ are
not assigned halo masses, since they will live in halos below the
$10^{11.635} \Msol$ halo mass limit.  The vast majority of such galaxies
are fainter than our $\MB = -19.0$ cut.

\begin{figure*}
  \epsscale{0.7}
  \plotone{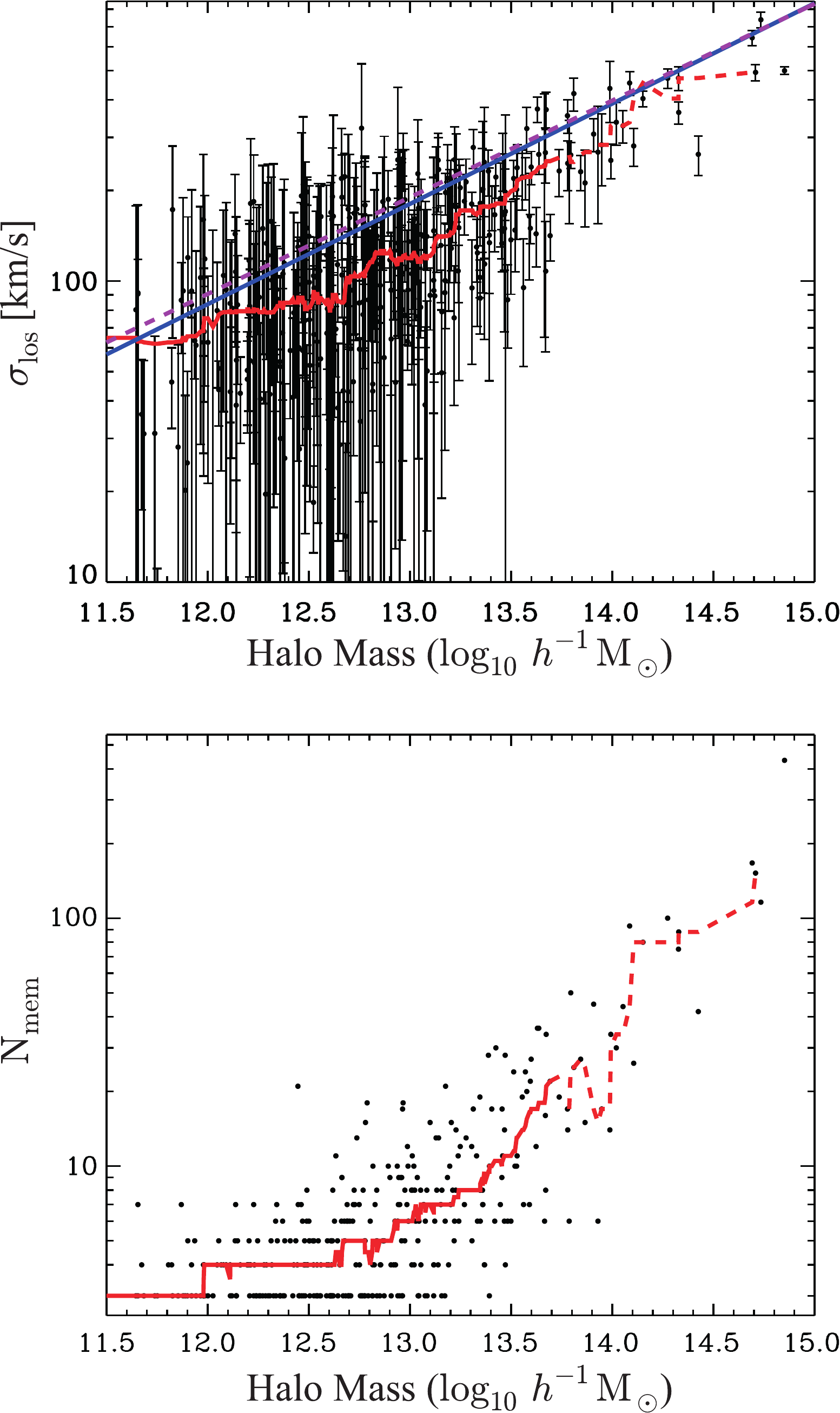}
  \caption{Y07-calibrated group halo mass versus group rest-frame
    velocity dispersion (top) and number of group members (below) for
    groups with at least 3 members. The red line indicates the
    median of the distribution as a function of halo mass, computed
    over a running bin of 50 galaxies (5 for the 25 highest mass
    groups, dotted line). The blue solid line in the top plot is the
    result of a simple virial prediction, and the magenta dashed line
    is the simple fitting function used by Y07.}
  \label{figure:masssigma}
\end{figure*}

Figure~\ref{figure:masssigma} examines how halo mass estimates for
groups with $\geq 3$ members correlate with line-of-sight velocity
dispersion $\sigma_{\rm los}$ (top) and number of confirmed members
$N_{\rm mem}$ (bottom).  $\sigma_{\rm los}$ is computed using the ``Gapper''
algorithm, appropriate for groups with a small number of members
\citep{Beers90}.  The overplotted median $\Mhalo$-$\sigma_{\rm los}$ relations
(using a running bin of 50 galaxies -- or 5 galaxies for the most
massive 25 groups) shows that a typical group halo mass of $\sim 10^{13}
\, h^{-1} \Msol$ corresponds to a measured velocity dispersion of
$\sigma_{\rm los} \sim 135 \kms$ and $N_{\rm mem} \sim 3$--18 (median $\sim 6$).
These masses are $\sim 40\%$ less than a simple prediction derived from
the virial relation $M_{dyn} = (3/G) \, R_{200} \, \sigma_{\rm los}^2$ with $R_{200}
= \sqrt{3} \, \sigma_{\rm los}/(10 \, H_0)$ (blue solid line), or the equivalent
relation used by Y07, $\sigma_{\rm los} = 397.9\kms \, (\frac{M_{halo}}{10^{14}
\, h^{-1} \Msol})^{0.3214}$ (magenta solid line). However, this is
expected; application of the group-finder algorithm to mock catalogs
predicts this $\sim 40\%$ discrepancy which results from the trimming of
high peculiar velocity members at the tails of the velocity histogram
(Y07).  Thus $\Mhalo \sim 10^{13} \, h^{-1} \Msol$ corresponds to a true
velocity dispersion of $\sigma_{\rm los} \sim 190\kms$, and galaxies with
large peculiar velocities will typically be assigned their own halo.

Y07 also classify galaxies as either {\it central} or {\it satellite}.
In reality, a ``central'' galaxy is just the most massive galaxy in any
given halo (i.e. equivalent to the Brightest Group Galaxy, BGG),
regardless of its actual physical location within the group, and all
other galaxies in the group are ``satellites''.  \citet{Skibba10} show
that a high fraction of so-called ``central'' galaxies are not in fact
located at the center of the group in terms of either projected position
or velocity: going from $\sim 25$\% for $10^{12} \, h^{-1} \Msol \leq
\Mhalo \leq 10^{13} \, h^{-1} \Msol$ to $\sim 40$\% for $\Mhalo \gtrsim
10^{13.7} \, h^{-1} \Msol$.  This must be a consequence of the dynamical
state of many of these systems which will be caught in the midst of a
major halo merger, or at a time before relaxation of the descendent
halo. Nonetheless, there are good theoretical and observational reasons
to treat the most massive galaxy differently from the other galaxies: it
should have the deepest potential well of any galaxy in the halo,
especially if it does sit at the center of the global potential (when
one exists after relaxation).  This means halo gas can cool more easily
onto this galaxy than it can onto any of the others. Observationally,
these two types of galaxy have been shown to behave quite differently on
average: central galaxies are more likely than satellites to be forming
stars and/or hosting radio AGN at fixed stellar and halo mass
\citep{Weinmann06,VonderLinden07,Best07,VonderLinden10}.  Therefore, we
use these classifications to examine the dependence of galaxy morphology
on whether the galaxy is a central or satellite galaxy as well as on the
underlying halo mass.

Of the 1340 galaxies in our SDSS-RC3 sample, 1194 have $B \leq 16$; the sample
size shrinks to 1064 when our $\MB \leq -19$ cut is also applied.  Within this
subsample, there are 911 galaxies which have halo masses assigned by Y07, with
729 of these being central galaxies and 182 being satellites. It is important to
note that our SDSS-RC3 sample does not include the richest, most massive galaxy
clusters: the maximum halo mass is $\sim 7 \times 10^{14} \Msol$, while the
maximum velocity dispersion is $\sim 740 \kms$.

\subsection{Stellar Masses}\label{sec:mstar}

To examine how the galaxy population varies with stellar mass, we
calculated stellar masses for each galaxy using the color-based
mass-to-light ($M/L$) ratios of \citet[][hereafter Z09]{Zibetti09},
using SDSS $g - i$ colors and $i$-band absolute magnitudes (including
the necessary k-corrections).  We prefer this approach over using the
stellar masses of \citet{Yang07}, which are based on the $M/L$ ratios of
\citet{Bell03}, because the Zibetti et al.\ $M/L$ ratios include the
effects of dust reddening and extinction, along with a spread of
possible star-formation histories.

We also investigated using the stellar mass derivations of
\citet[][hereafter G05]{Gallazzi05}\footnote{These stellar masses, as
well as H$\alpha$ linewidth measurements, were taken from the MPA public
webpage http://www.mpa-garching.mpg.de/SDSS/DR4/}, which take advantage
of SDSS spectroscopy for each galaxy.  The G05 masses were
estimated by fitting five spectral absorption features to estimate a
$z$-band $M/L$ ratio, applied to the $z$-band luminosity. The
drawback of this approach, from our perspective, is that the SDSS fiber
spectra sample a relatively small, central region of the galaxies: since
RC3 galaxies usually have diameters $\gtrsim 1\arcmin$, the 3-arcsec
SDSS fiber aperture captures only the inner region of the galaxy. For
spiral galaxies, this can mean that the spectroscopic $M/L$ reflects the
bulge or nuclear region, rather than the galaxy as a whole. In addition,
spectroscopy-based stellar masses are not available for some galaxies
(142 of the 1340 galaxies in our full RC3-SDSS matched sample).

In the rest of this paper, we mainly concentrate on the Z09-based
stellar masses, but we tested all trends with stellar mass against the
G05 and Y07 masses as well, and note when these provided different
results.

\subsection{Systematics and Limits in Stellar and Halo
Masses}\label{sec:masslimits}

\begin{figure*}
  \epsscale{1.00}
  \plotone{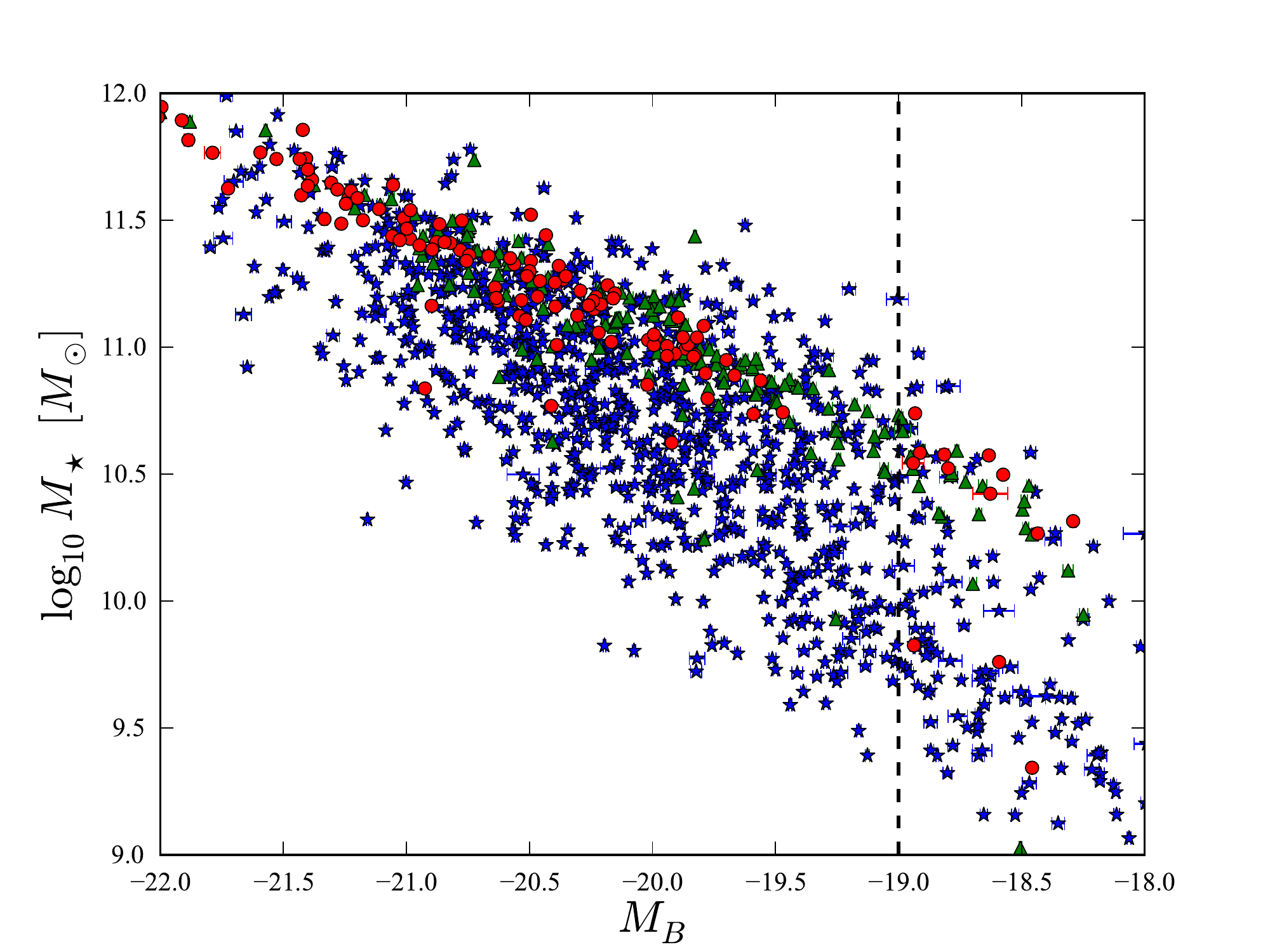}
  \caption{Stellar mass ($\Mstellar$) versus $B$-band luminosity
    ($\MB$) for all galaxies in our sample. Elliptical, S0, and spiral
    galaxies are indicated by red circles, green triangles, and blue
    stars, respectively. The vertical dashed line marks our $\MB =
    -19$ luminosity cutoff.}
  \label{figure:mbmstellar}
\end{figure*}

Figure~\ref{figure:mbmstellar} illustrates how galaxies of different
morphological type populate the $\MB$-$\Mstellar$ plane. (The $B$-band
luminosity $\MB$ is computed applying both distance modulii in our
chosen cosmology and k-corrections based on the {\it kcorrect} code of
\nocite{Blanton07kcor}Blanton \& Roweis 2007.)  Spiral galaxies tend to
have lower $B$-band stellar $M/L$ light ratios, although the
highest $M/L$ ratio galaxies are also spirals (with
dust-obscured $B$-band light). Ellipticals and S0s populate a fairly
tight relation, indicating that there is little difference between a
luminosity-based cut and a mass-based cut for E versus S0
classifications.  At our selected luminosity cut of $\MB = -19.0$, an
early-type galaxy has a typical stellar mass of $\Mstellar \sim
10^{10.5} \Msol$.  Cutting in mass instead of luminosity would reduce
the unweighted total spiral fraction by less than $4\%$.

Figure~\ref{figure:mbmstellar} also shows that spiral galaxies with $\MB
< -19$ can have masses down to $\sim 10^{9.5} \Msol$.  Because the halo
mass estimates are ultimately based on galaxy stellar masses, these different
stellar-mass limits have implications for our ability to find different
galaxy types as a function of halo mass and galaxy status (central
versus satellite). The simplest case is for central galaxies. If we
consider isolated galaxies (where the galaxy is automatically the
central galaxy of its halo), then a stellar-mass limit of $10^{10.5}
\Msol$ implies a halo mass limit of $\sim 10^{11.8} \, h^{-1} \Msol$ for
central E/S0 galaxies. The lower mass limit for spiral galaxies means
that they can be found as central galaxies for halos with lower masses,
down to the Y07 completeness limit ($\Mhalo = 10^{11.63} \, h^{-1} 
\Msol$). This means that for halos masses $< 10^{12} \, h^{-1} \Msol$,
we should expect to find few or no central E/S0 galaxies, \textit{solely
due to our luminosity cut}.

A similar, albeit more complicated, effect applies to satellite
galaxies. The minimum halo mass for a satellite galaxy can be estimated
assuming that it is the second most massive galaxy in a two-galaxy
group, with the central galaxy only marginally more massive. In this
case, a $10^{10.5} \Msol$ \textit{satellite} galaxy will have a halo
mass $\gtrsim 10^{12.5} \, h^{-1} \Msol$, and we should expect to
undercount E/S0 satellite galaxies in less massive halos. A spiral
satellite with $\Mstellar \sim 10^{9.5} \Msol$, on the other hand, could
reside in a halo with total stellar mass content $\Mstellar \gtrsim
10^{9.8} \Msol$ corresponding to halos of low mass (potentially down to
the Y07 completeness limit).

The fact that our $\MB < -19$ cut largely excludes E and S0 galaxies with
stellar masses $\lesssim 10^{10.5} \Msol$, while still including a large number
of spiral galaxies with smaller stellar masses, means that direct comparisons of
E, S0, and spiral galaxies will be biased for stellar masses $< 10^{10.5}
\Msol$. Consequently, when we consider fractions of different morphological
types as a function of stellar or halo mass (e.g., in
Sections~\ref{sec:lumdep} and \ref{sec:morphcensat}), we limit ourselves to
$\Mstellar > 10^{10.5} \Msol$. This limit does not apply when we consider
fraction \textit{within} a given morphological class -- e.g., the fraction of
ellipticals with optical AGN spectra, or fractions of spirals with outer rings.

\section{Results}\label{sec:results}

In this section we look at how various galaxy classifications -- broad
Hubble types (elliptical, lenticular, spiral), spectroscopic properties,
and morphological substructure (bars and rings) -- depend on global
galaxy properties and on group properties.

\subsection{Uncertainties and Significance Testing}

We use fractions to discuss the morphological composition of the
local galaxy population. Fractions are by definition relative quantities
and as such allow comparisons between different types, doing away with
the need to renormalize to the total galaxy population in a given bin.
However, the reader should keep in mind that focusing on fractions does
have its own problems.  For example, the total galaxy population in a
given bin of stellar or halo mass, or luminosity, is {\it not}
necessarily conserved with time. Since we consider multiple
morphological types (e.g., elliptical versus S0 versus spiral), the
fraction of a given type can change due to transformations between other
types, and not just because the total number density of the first type
is increasing or decreasing (e.g., if the S0 number density stays
constant but spirals merge to form ellipticals, then the total number of
E+Sp galaxies decreases and the S0 \textit{fraction} will increase).

Our plots show fractions for each specified type within various bins.
The error bars in these plots are 68\% confidence limits from the
\citet{Wilson27} binomial confidence interval, which is a more accurate
way of estimating binomial uncertainties than the commonly used Gaussian
approximation, especially as the frequencies approach 0 or 1; see
\citet{Brown01} for a comprehensive discussion of binomial
uncertainties.

In the case of \textit{weighted} counts, there is less guidance on the
proper way to compute uncertainties. We estimate the uncertainties by
first rescaling all (weighted) counts so that the total counts are equal
to the original (unweighted) total counts in a given bin, and then
computing the Wilson confidence limits using the rescaled counts.

We estimate the \textit{significance} of apparent trends in these plots
via a logistic regression analysis. In logistic regression, the
probability of a binomial property (e.g., galaxy is S0 or not, galaxy is
barred or not) as a function of some independent variable $x$ is modeled
using the following function: \begin{equation} P \, = \, \frac{1}{1 +
e^{-\alpha + \beta x}} \end{equation} where $P$ is the probability for a
galaxy having the particular property. The coefficient $\alpha$
corresponds approximately to an intercept value, while $\beta$ is
analogous to a slope and measures how strong the trend is. Note that the
function always has values between 0 and 1 (appropriate for a
probability), and is either monotonically increasing or decreasing.

In principle, we could fit straight lines to the binned frequencies and
use the null probability for a nonzero slope as an estimate of a trend's
significance, but the logistic approach has two key advantages. First, a
linear fit will yield (meaningless) frequency values above 1 or below 0
at some point, while the logistic curve is bounded between 0 and 1.
Second and perhaps more importantly, linear fits to frequencies will be
biased by the specific binning scheme used (number and spacing of bins),
and by the assumption of Gaussian errors -- an assumption that is not
necessarily true for frequencies, especially for frequencies close to 0
or 1. Logistic regression uses all data points individually, making
binning and error assumptions irrelevant.

The specific logistic regression code we use is from the Survey
package\footnote{http://faculty.washington.edu/tlumley/survey/}, implemented in
the R statistical language\footnote{http://www.r-project.org/}; this package
allows for individual data points to have weights. The fitting process yields
null-hypothesis probabilities for the intercept $\alpha$ and slope $\beta$
(i.e., the probability that a logistic model with a value of 0 for the given
coefficient could explain the data); we use the null-hypothesis probability for
$\beta$ ($\Plog$) as an estimate of an apparent trend's significance. We
also quote the best-fit value of $\beta$ and its uncertainty; positive values
mean the frequency increases with the independent variable, while negative
values indicate a trend of decreasing frequency. We caution the reader that
this is \textit{not} a universal indicator for a trend: trends more complicated
than monotonic, smooth increase or decrease in probability could be poorly fit
by the logistic model. Nonetheless, we feel this is superior to the commonly
used approach of least squares linear fitting of binned data, for the reasons
given above. A two-sample test such as the Kolmogorov-Smirnov test is also not
advisable, both because it is calibrated only for unweighted data and because it
measures the maximum, local difference between two samples (e.g. ellipticals and
non-ellipticals) rather than the significance of a global trend.

\subsection{Morphological Fractions Versus Luminosity and Stellar
Mass}\label{sec:lumdep}

\begin{figure*}
  \epsscale{1.00}
  \plotone{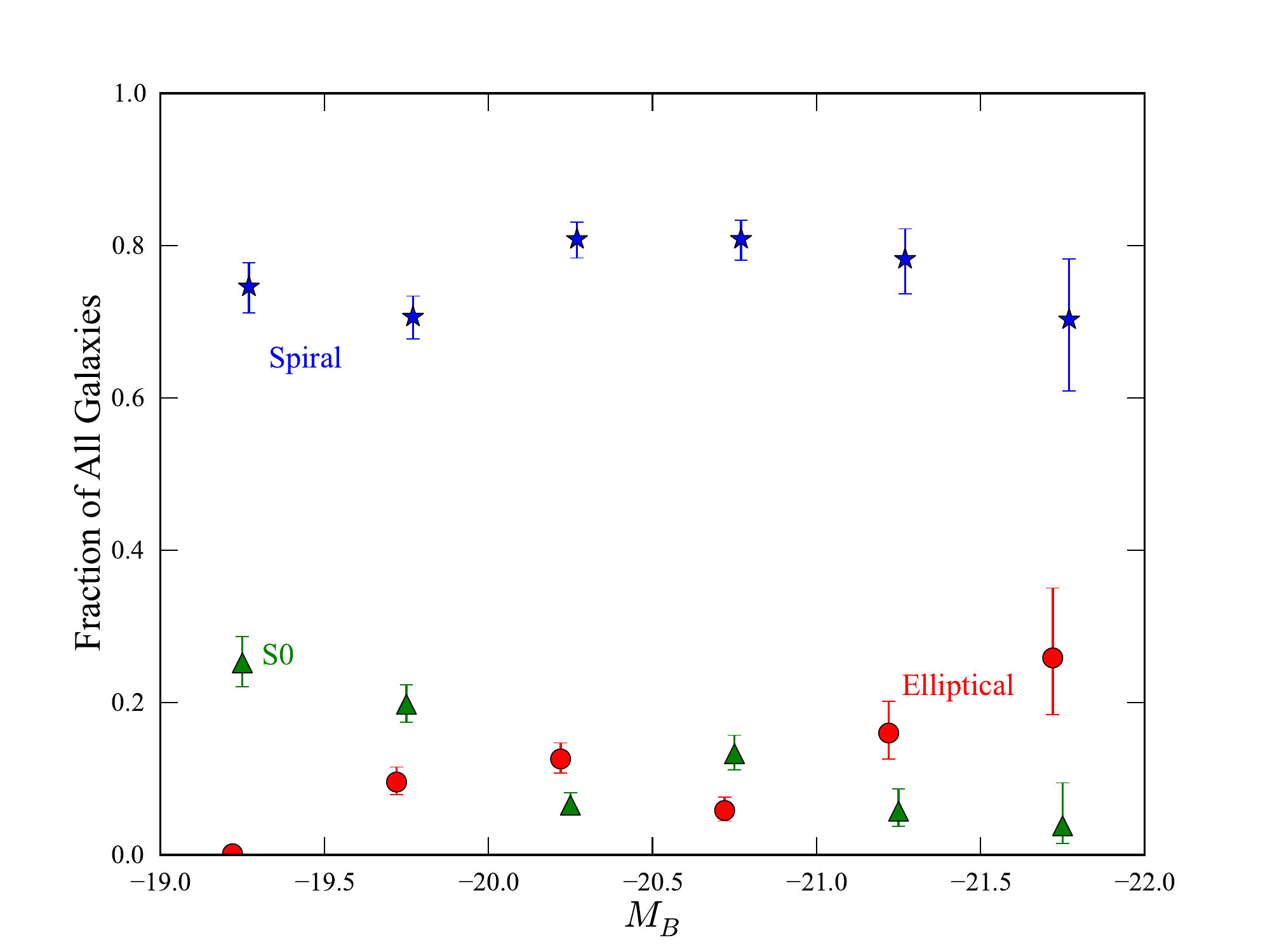}
  \caption{Fractions of elliptical (red circles), lenticular (green
    triangles), and spiral (blue stars) galaxies as a function of {\it
      B}-band luminosity, with selection weights applied. Binomial
    errors are computed using the \citet{Wilson27} method.}
  \label{figure:ftypemb}
\end{figure*}

Figure~\ref{figure:ftypemb} shows how the fractions of elliptical,
lenticular, and spiral type galaxies depends upon {\it B}-band
luminosity within the full SDSS-RC3 sample.  Selection weights are
applied to galaxies; when plotting against luminosity, it is not
necessary to apply $V/V_{\rm max}$ weights.

The overall elliptical fraction is low, but clearly increases with luminosity
($\beta = -1 \pm 0.24, \Plog = 1.3 \times 10^{-5}$; note that this indicates a
frequency that decreases as $\MB$ becomes more positive).  The observed
fraction of S0 galaxies {\it decreases} with luminosity ($\beta = 0.8 \pm
0.3, \Plog = 0.0088$), whilst the fraction of spiral galaxies show little, if
any dependence on luminosity down to $\MB \sim -19$ ($\beta = -0.12 \pm
0.22, \Plog = 0.59$). From a theoretical standpoint, this is perhaps
surprising, indicating that the overall probability that spiral arms have faded
(likely associated with the suppression of star formation in a disk galaxy) is
not a strong function of its luminosity.

\begin{figure*}
  \epsscale{1.00}
   \plotone{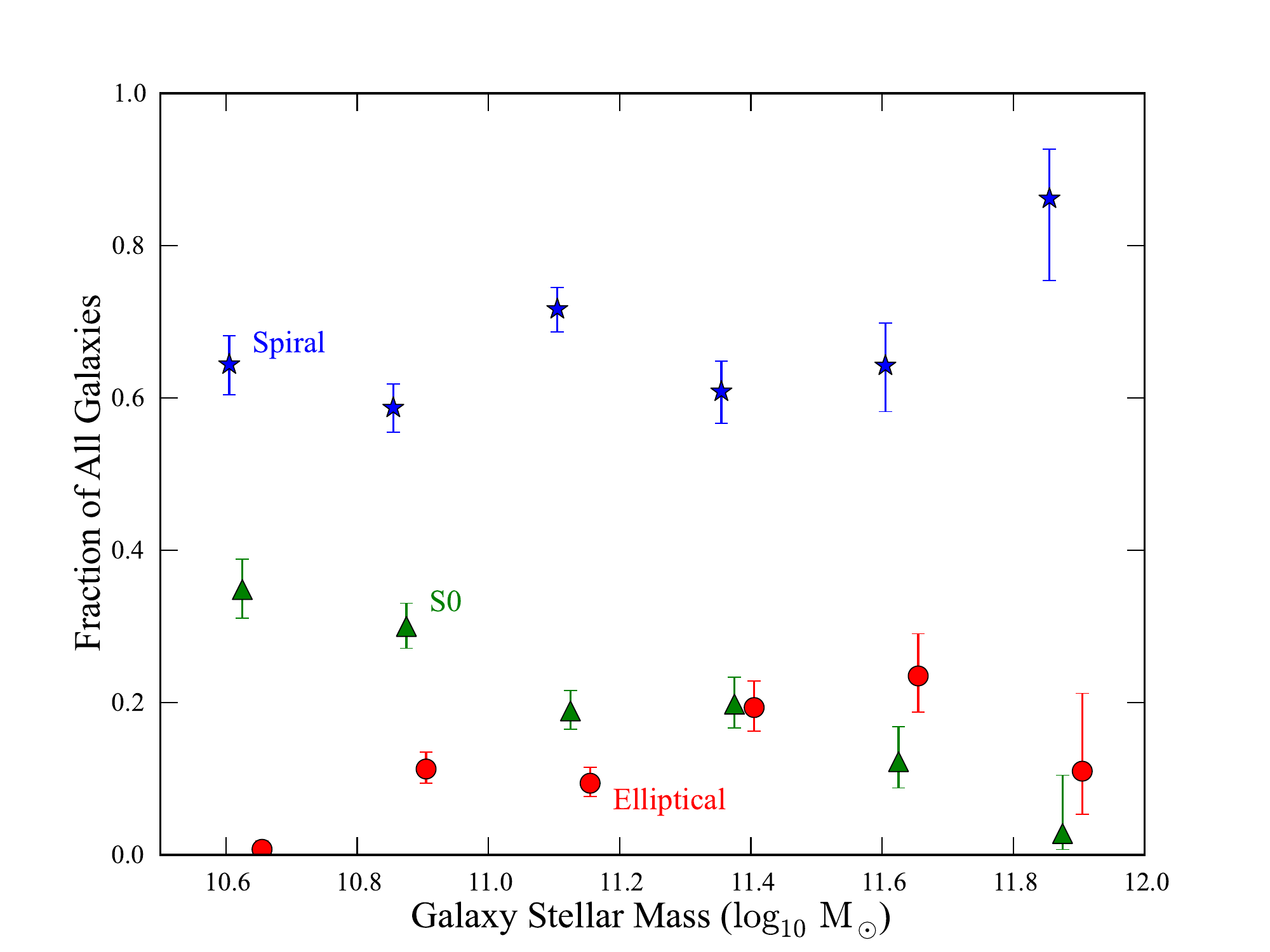}
  \caption{Fractions of each morphological type (elliptical, lenticular
    (S0), and spiral) for $\MB < -19$ galaxies as a function of stellar mass, 
    weighted using selection and $V/V_{\rm max}$ weights.}
  \label{figure:ftypemstar}
\end{figure*}

We examine the stellar mass dependence of morphological fractions in
the top panel of Figure~\ref{figure:ftypemstar}, cut at $\MB = -19$
and weighted to account for both selection and volume ($V_{\rm
  max}$). We also restrict our plots, and our logistic regression
analysis, to galaxies with $\Mstellar \geq 10^{10.5} \Msol$, because
selection effects can lead to artificial suppression of elliptical and
S0 populations relative to spirals below this mass limit
(Section~\ref{sec:masslimits}). The fraction of ellipticals does
increase significantly with increasing stellar mass ($\beta = 3.2 \pm
0.64, \Plog = 4.3 \times 10^{-7}$).  Overall, the S0 fraction appears
to decrease as stellar mass increases, although this is not formally
significant ($\beta = -0.32 \pm 0.74, \Plog = 0.66$). The fraction of
spiral galaxies is essentially constant at $\sim 65$\% over this mass
range ($\beta = -0.74 \pm 0.68, \Plog = 0.27$), similar to what was
seen for trends with \MB.

\subsection{Morphological Fractions Versus Halo Mass}

\begin{figure*}
  \epsscale{0.65}
  \plotone{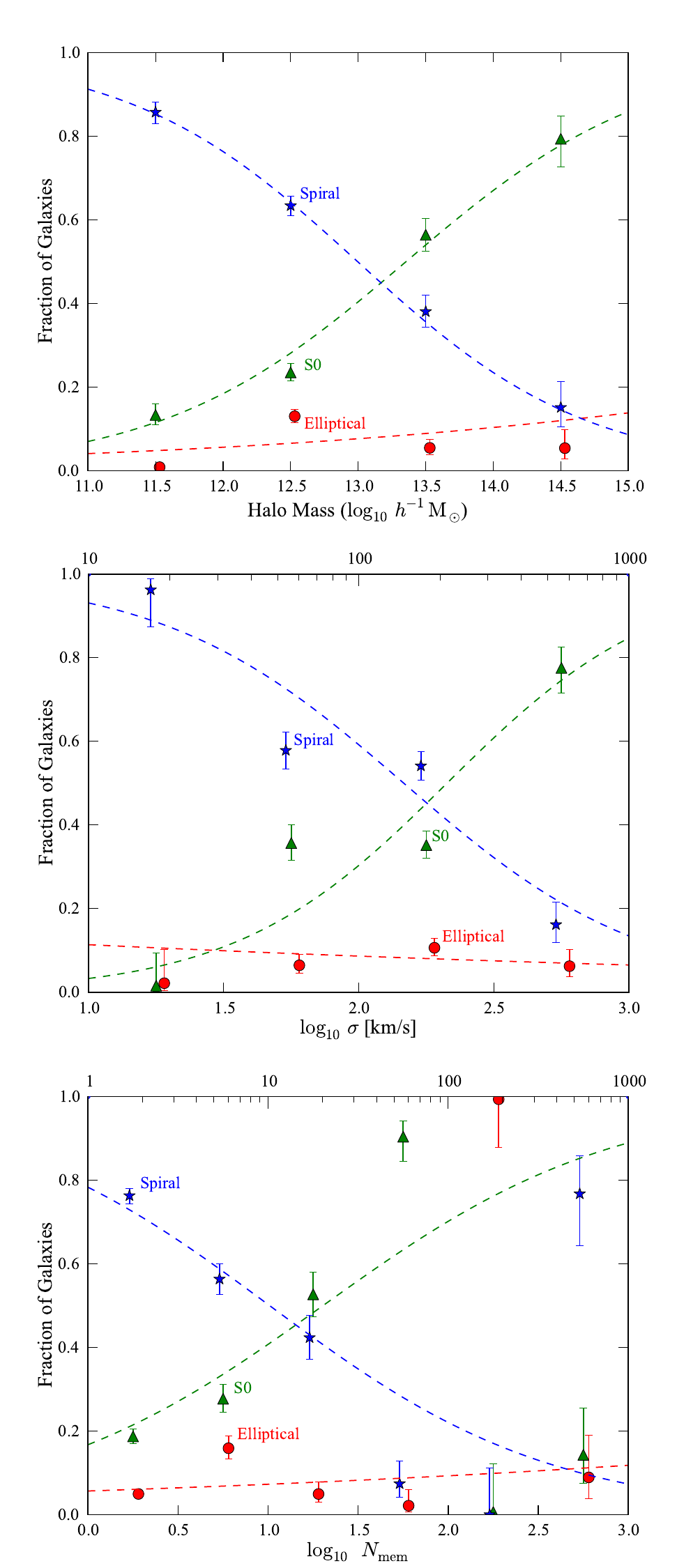}
  \caption{As for Figure~\ref{figure:ftypemstar}, but now showing morphological 
  fractions as a function of halo mass (top), group velocity dispersion (middle),
  and number of group members (bottom); only galaxies with $\Mstellar > 10^{10.5} \Msol$
  are considered.  Dashed lines indicate logistic fits to the (unbinned) data. }
  \label{figure:ftype-combined}
\end{figure*}

We present the fraction of galaxies with $\MB \leq -19.0$ and
$\Mstellar > 10^{10.5} \Msol$ as a function of group/halo properties
in Figure~\ref{figure:ftype-combined}. The top panel presents the
trends as a function of halo mass.  This shows a strong dependence of
morphological type on halo mass.  The elliptical fraction is
practically zero below $\Mhalo \sim 10^{12} \, h^{-1} \Msol$ (but this
is most likely due to the selection effect discussed in
section~\ref{sec:masslimits}), then rises to a roughly constant level
of $\sim 5$--10\% for more massive halos.  The lenticular fraction
increases dramatically from $\sim 10$\% in the very lowest halo masses
to $\sim 80$\% at the high-mass end (the dashed green line shows the
logistic fit, which has $\beta = 1.1 \pm 0.32, \Plog = 0.00071$. The
fraction of spiral galaxies naturally compensates for these trends
with halo mass, decreasing from $\sim 85$\% to $\sim 15$\% going from
the lowest mass to the highest mass halos ($\beta = -1.2 \pm 0.31,
\Plog = 0.00013$ for the logistic fit shown by the dashed blue
line). Significant trends for S0s and spirals are also found with
group velocity dispersion $\sigma_{\rm los}$ and number of members
$N_{\rm mem}$ as the independent variable (logistic fits, shown in the
center and lower panels). 

The most dramatic outlying point in Figure~\ref{figure:ftype-combined}
is the highest bin of $N_{\rm mem}$ (bottom panel), in which the spiral
fraction jumps to almost $80\%$. This bin is populated by a single group, the
Abell 2199 supercluster \citep[see e.g.][]{Rines02}, which has
apparently been merged by the \citet{Yang07} group-finder algorithm into
a single, overmassive group with 433 members. Most of the 18 RC3
galaxies belonging to this system lie outside the collapsed regions of
the supercluster (including Abell 2197 and Abell 2199).  Their
spiral morphologies are therefore not surprising.

\subsection{Comparing Central and Satellite Galaxies}\label{sec:morphcensat}

\begin{figure*}
  \epsscale{1.22}
  \plotone{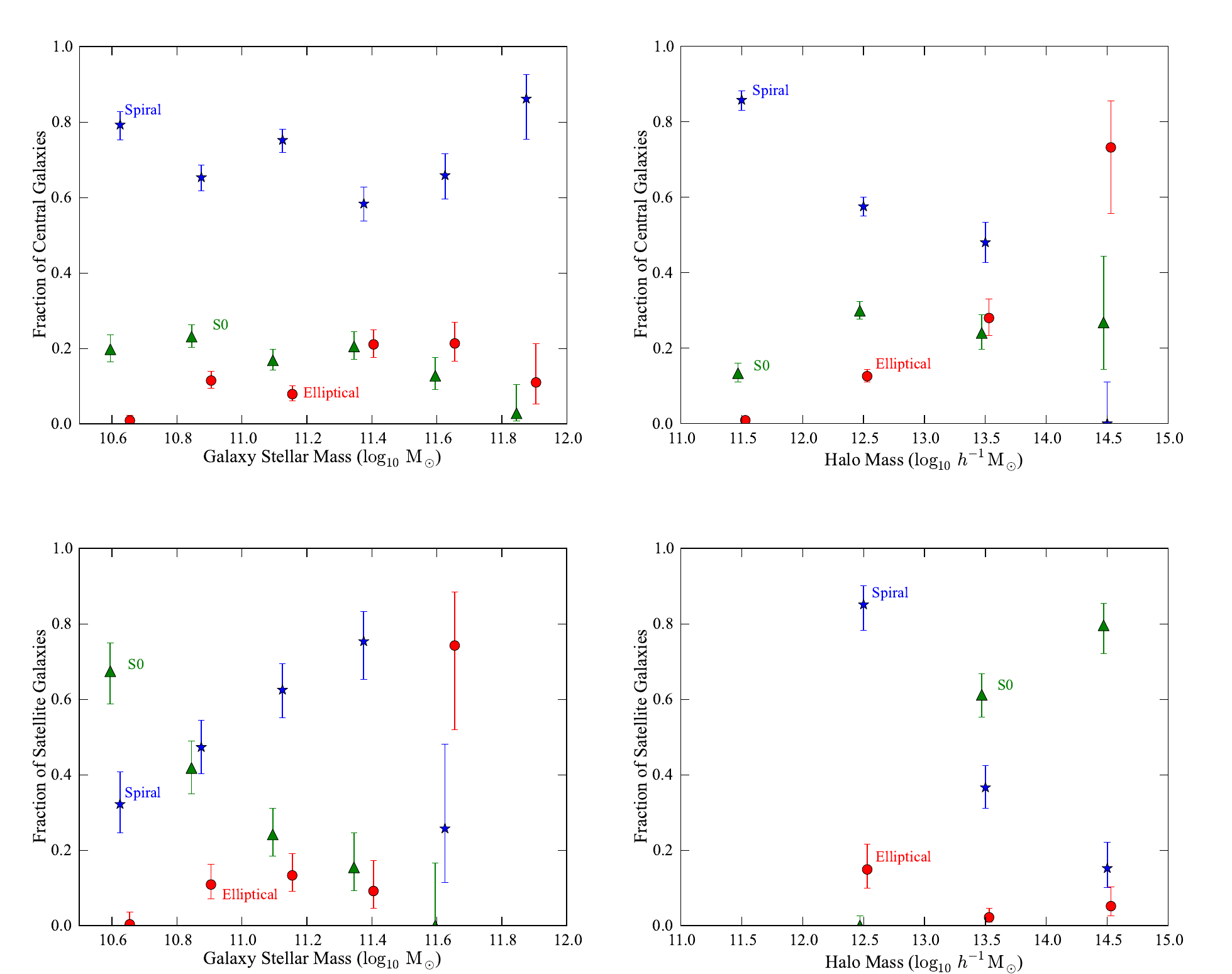}
  \caption{As for Figure~\ref{figure:ftypemstar}, but now showing morphological
  fractions versus $\Mstellar$ (left) and $\Mhalo$ (right) separately for central
  galaxies (top) and satellite galaxies (below); only galaxies with 
  $\Mstellar > 10^{10.5} \Msol$ are considered.}
  \label{figure:ftypemhalocensat}
\end{figure*}

Figure~\ref{figure:ftypemhalocensat} splits the galaxy population into
central and satellite galaxies, which (at least theoretically) should
be subject to different physical processes. The morphological
fractions are plotted against both the stellar mass of individual
galaxies ($\Mstellar$, left panels) and halo mass ($\Mhalo$, right
panels). Once again, we restrict the plots and logistic-regression
analyses to galaxies with $\Mstellar > 10^{10.5} \Msol$ (see
Section~\ref{sec:lumdep}).

The dependence of galaxy morphology on galaxy stellar mass shows some
interesting differences when we consider central and satellite
galaxies separately. The roughly constant fraction of spirals as a
function of stellar mass for all galaxies
(Figure~\ref{figure:ftypemstar}) is replicated by the central
galaxies; but for satellites we evidence of a trend where the
frequency of spirals \textit{increases} with mass, at least up to
$\Mstellar \sim 10^{11.5} \Msol$. For S0s, the apparent differences
are even stronger: central S0s are a roughly constant fraction at all
masses $\lesssim 10^{11.5} \Msol$, while satellite S0s show a steep
drop in frequency as stellar mass increases (however, this trend is
not formally signficant: $\beta = -3 \pm 1.8, \Plog = 0.1$). Clear
trends are harder to discern for ellipticals; in general, they
resemble the elliptical trend for all galaxies
(Figure~\ref{figure:ftypemstar}), with frequency increasing weakly but
significantly with stellar mass ($\beta = 3.1 \pm 0.64, \Plog = 1.2
\times 10^{-6}$ for central ellipticals).

When we turn to the question of how the morphological fractions of
central and satellite galaxies depend on \textit{halo} mass, we see
some striking differences. This is not actually true for spiral
galaxies, which decline in frequency as halo mass increases for both
central and satellite galaxies, just as we saw for galaxies in general
(Figure~\ref{figure:ftype-combined}). But when we look at elliptical
galaxies, we see a clear dichotomy: the elliptical fraction for
satellite galaxies is roughly constant ($\beta = -0.34 \pm 0.93, \Plog
= 0.72$), while the fraction for \textit{central} galaxies is a
steeply increasing function of halo mass ($\beta = 2.2 \pm 0.38, \Plog
= 8.2 \times 10^{-9}$).


S0 galaxies also show a dichotomy: the fraction of central galaxies
which are S0 is roughly constant (the apparent decrease for $\Mhalo <
10^{12} \, h^{-1} \Msol$ is likely a result of the selection effect
discussed in Section~\ref{sec:masslimits}, above), but the fraction of
\textit{satellite} galaxies which are S0 jumps from $0^{+2.6}_{-0}$\%
to $69.0^{+4.3}_{-4.6}$\% in higher-mass halos! This appears to be a
highly significant difference; the logistic-regression value of $\Plog
= 0.035$ probably understates the significance because an abrupt
transition like that seen here is not well modeled by the logistic
curve (it would, of course, not be well modeled by a simple linear fit
to the binned fractions, either).

One clear implication from this analysis is the importance of group halo
mass for determining galaxy morphology. For central galaxies, halo mass
is clearly more important than galaxy mass. The frequency of central spirals is
high and essentially constant over the stellar mass range $\Mstellar =
10^{10.5}$--$10^{12} \Msol$.  But when we look at variations in halo mass,
we can see central spirals being replaced by central ellipticals as the
halo mass grows; only the central S0 frequency seems to be independent
of halo mass.

Halo mass also has a strong effect on satellite galaxies, although in a
somewhat different fashion. As halo mass grows, the fraction of
satellites which are spirals falls, just as happens for central
galaxies. But satellite spirals in higher-mass halos are clearly being
replaced by \textit{lenticular} galaxies, which become the dominant type
of satellites for $\Mhalo > 10^{13} \, h^{-1} \Msol$.

\subsection{Co-evolution of Morphology, Star Formation and AGN
Activity}\label{sec:morphactivity}

Physical processes which transform spiral galaxies into elliptical or
lenticular galaxies should describe the destruction of a galaxy's disk
in the case of ellipticals, the enhancement of the bulge component for
both ellipticals and bulge-dominated lenticulars, and the dissolution of
spiral arms (lenticulars). Such processes should lead to the observed
correlations between morphology and environment.  In addition, these
processes may be responsible for circumnuclear starbursts and AGN, along
with the removal of gas and the general suppression of star formation.

We use the spectroscopic information from the SDSS survey to
characterize the ongoing star formation and nuclear activity in our
sample. H$\alpha$ emission-line flux correlates strongly with the star
formation rate, and is strongly bimodal for galaxies with and without
significant star formation or nuclear activity \citep{Balogh04}.  SDSS
spectral fibers sample only the central $3\arcsec$ (diameter) of each
galaxy, corresponding to 0.61--2.35 kpc at $z = 0.01$--0.04.  This
undersamples the galaxy as a whole, and will miss much of the ongoing
star formation and H$\alpha$ emission at large radii.

We use the \citet[][hereafter B04]{Brinchmann04} calibration of
H$\alpha$ equivalent width, which corrects the emission flux for
underlying stellar absorption. We find a bimodality in this quantity,
with a {\it low emission} peak at $-0.5$\AA\ (emission is negative),
which we presume should be at $0$\AA\ (no emission for a truly passive
galaxy).  We define a galaxy as ``passive core'' if the H$\alpha$
equivalent width is within $2.5\sigma$ of this peak (where $\sigma$ in
this case is the error in equivalent width, as estimated by B04, scaled
up by a factor 2.473 calibrated to repeat measurements of the same
galaxy)\footnote{http://www.mpa-garching.mpg.de/SDSS/DR4/raw\_data.html}
. The typical value of 2.5$\sigma$ is $\sim 0.6$\AA\ for these galaxies.

\begin{figure*}
  \epsscale{0.7}
  \plotone{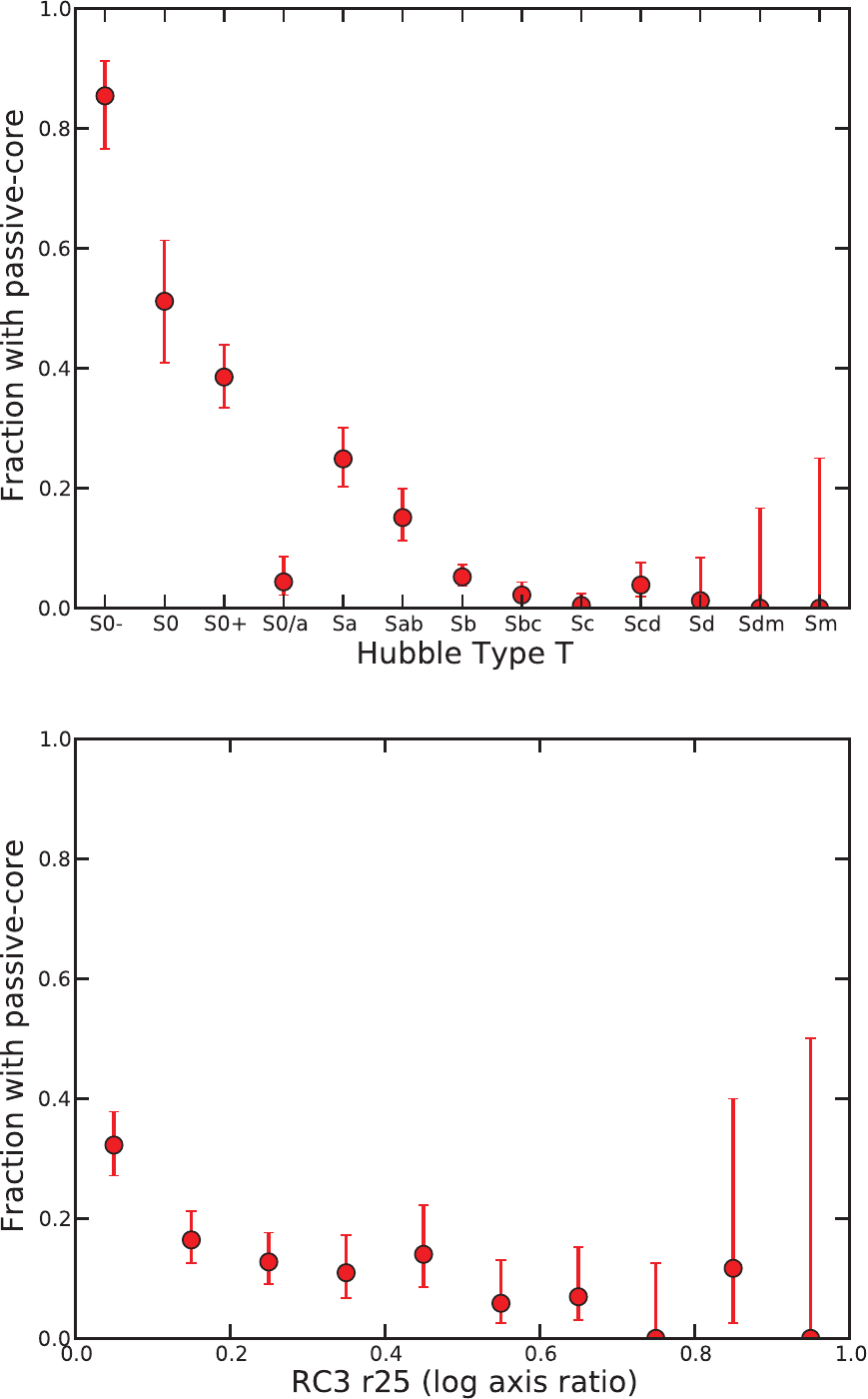}
  \caption{The fraction of disk galaxies with a passive core (SDSS
    fiber spectra lacking significant H$\alpha$ emission, see text) versus
    Hubble type (top) and, restricted to S0/a--Sb types, versus log
    axis ratio $a/b$ (bottom). A high fraction of S0s have passive
    cores, as do some S0/a--Sb type spirals (but very few later
    types). Most passive-core spiral galaxies are close to face-on
    (low axis ratios).}
  \label{figure:fSFSphtar}
\end{figure*}

The upper panel of Figure~\ref{figure:fSFSphtar} shows the fraction of
disk galaxies with a passive core as a function of Hubble type.
Passive-core galaxies are, unsurprisingly, most common amongst S0
galaxies, and the fraction decreases going from the RC3 type S$0^{-}$
through S0 to S$0^{+}$, going from E-like to spiral-like S0s.  Among
spirals, Sa galaxies are much more likely to have passive cores than
later Hubble types ($\sim 25\%$ Sa, $\sim 15\%$ Sab, $\sim 5\%$ Sb,
going to zero for later types).  The anomalously low S0/a passive-core
fraction is difficult to explain, and we avoid interpreting this for
now.

The lower panel of Figure~\ref{figure:fSFSphtar} shows the fraction of
S0/a--Sb galaxies with a passive core as a function of the RC3 value r25.
This is the logarithmic axial ratio $\log_{10} a/b$, where $a$ and $b$
are the semi-major and semi-minor axes measured out to a surface
brightness of 25 mag arcsec$^{-2}$ in the $B$-band. Low values of r25 imply
near circular isophotes.  The core of an early-type spiral is much more
likely to be passive for a face-on inclination (low axis ratio), for
which there will be no fiber contribution from the outer disk.  In such
early-type, face-on spirals the SDSS fiber spectrum can be dominated by
bulge light.  These correlations therefore suggest that star formation
in the disk can be either truncated or heavily obscured in the inner
regions, so that no significant H$\alpha$ emission is detected in the
fiber.  Since we find a very low passive fraction in highly inclined,
early-type spirals (where the outer disk is likely to be projected into
the SDSS fiber aperture), we infer that the outer disks of early-type
spirals are typically still forming stars, and that most passive-core
spirals are passive only in the inner regions.  This is consistent with
visual inspection of color JPEG images of these galaxies: in many cases
the inner bulge+bar region appears red, with blue outer spiral features,
often separated by a ring.

The resulting picture of an inner truncation for star formation is a
necessary simplification, given the limitations of our data. For
comparison, longslit spectroscopy of four spiral galaxies lacking
emission in the SDSS fiber revealed emission in the outer regions of two
-- and strong Balmer line absorption indicated recent star formation in
the other two \citep{Ishigaki07}.

\begin{figure*}
  \epsscale{1.00}
  \plotone{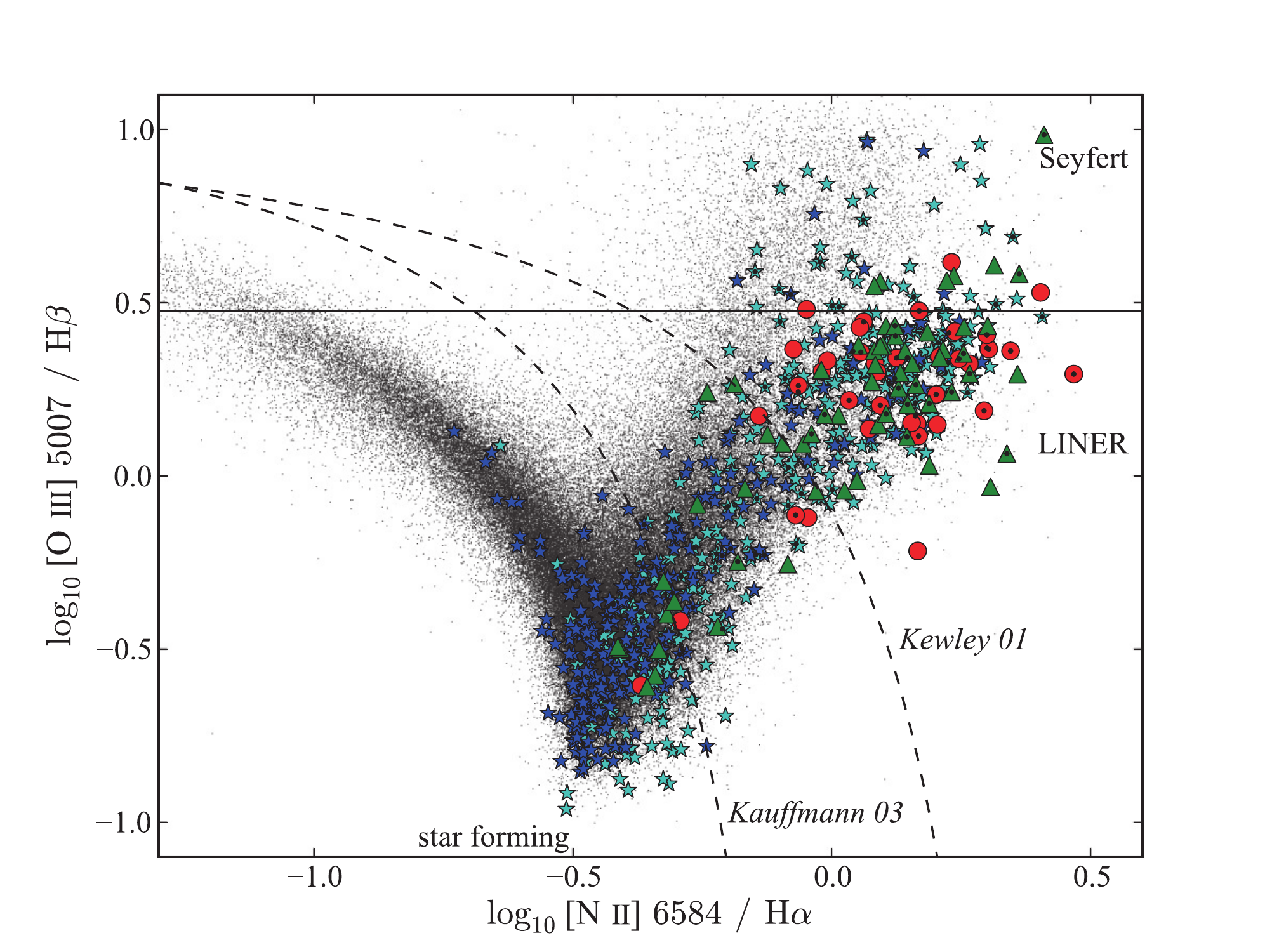}
\caption{ The ``BPT'' (Baldwin et al, 1981) emission-line ratio
  diagnostic diagram, showing the full parent sample (small points, all
  galaxies with S/N(all lines) $> 3$) overplotted with galaxies from our
  morphologically classified sample (galaxies with S/N(all
  lines) $> 2$), including elliptical (red circles), S0 (green
  triangles), S0/a-Sb (cyan stars), and Sc-Sm (blue stars)
  galaxies. Symbols with black dots at the centre contain a ``radio-AGN''
  source (see text for formal definition). Most elliptical and S0 galaxies, and many early-type
  spirals, have AGN-like line ratios (lying to the right of the
  \citet{Kauffmann03} and \citet{Kewley01} lines), indicating hard
  ionization fields inconsistent with starbursts, but with
  [\ion{O}{3}]/H$\beta$ ratios typically lower than than 3, suggestive of
  LINERs rather than Seyfert galaxies. Recall that the parent sample
  goes to higher redshift, which means that SDSS fibers sample light
  out to larger physical radii.}
  \label{figure:bpt}
\end{figure*}

Star formation is not the only possible source of H$\alpha$ emission, of
course. H$\alpha$ emission traces ionized gas -- and therefore requires
both gas and ionizing radiation.  The nature of that ionizing radiation
can be explored via emission-line ratios, tracing the relative
importance of different ionization levels and subsequent transitions. In
Figure~\ref{figure:bpt} we use the [\ion{N}{2}]$\lambda6584/$H$\alpha$
vs [\ion{O}{3}]$\lambda5007$/H$\beta$ diagnostic line-ratio diagram,
commonly used to separate normal star forming galaxies from harder
radiation fields typical of Seyfert and LINER (low-ionization nuclear
emission-line region) type galaxies \citep[][the ``BPT''
diagram]{Baldwin81}.  The small points in the background trace the
parent sample, demonstrating the overall distribution of galaxy
line-ratios (where S/N(all lines) $> 3$). Overplotted are galaxies in
the RC3-SDSS sample with S/N(all lines) $> 2$, keyed by morphology (see
caption).  Anything to the right of the \citet{Kewley01} dashed line
cannot be explained by normal starburst models, whilst the
\citet{Kauffmann03} dashed line demarcates the boundary of normal star
forming galaxies (to the left). It should be noted that RC3-SDSS
galaxies are at lower redshifts than the typical parent sample galaxy,
and so fiber spectra will be more dominated by nuclear emission.
Nonetheless, it is interesting that almost all E/S0 galaxies with
emission are classified as LINERs ([\ion{O}{3}]$/$H$\beta < 3$ implies a
softer-than-Seyfert ionization field). Some S0s extend to the region
between the two dashed lines \citep[composite/transition
systems,][]{Kauffmann03}, whilst S0/a--Sb (and some later-type) spirals
extend from the bottom (high metallicity) end of the star forming locus,
up through the composite and LINER region to the Seyfert regime (with
LINERs again the dominant population).  As we sample only the central
$\lesssim 1\kpc$, the ionization source may well be related to accretion
onto a super-massive black hole (SMBH).

We classify galaxies with emission lines into two basic categories.
Anything with broad emission lines or lying to the right of the
\citet{Kewley01} line is termed ``AGN''. All other emission line
galaxies are called ``star-forming''.  Whilst these are commonly
used definitions, we note that LINER-like ionization may also result
from older stellar populations, shocks or interaction with hot, X-ray
emitting gas \citep[see e.g.][]{Sarzi10,Capetti2011} and not solely
from accretion onto a SMBH.

To understand the possible role of AGN feedback for the suppression of star
formation, it is also useful to examine the radio properties of these galaxies.
Models of galaxy formation invoke {\it radio-mode} AGN feedback, which
suppresses cooling onto galaxies living at the centers of massive halos by
coupling the kinetic energy of a radio jet to the cooling gas
\citep[e.g.][]{Bower06,Croton06}. To examine the role of radio-mode feedback, we
cross-correlated our sample with the VLA (Very Large Array) FIRST (Faint Images
of the Radio Sky at Twenty-Centimeters) survey \citep{FIRST}.  This 1.4GHz
survey has a nominal detection threshold of 1 mJy, with a 90\% confidence
positional error circle of radius 1\arcsec{} (0.5\arcsec{} at the 3 mJy level).
We first identified ``nuclear'' sources by requiring a match between SDSS and
FIRST positions within 2\arcsec; this yielded 261 FIRST sources with fluxes $>
1$ mJy matched to the SDSS-RC3 sample.

Continuum radio emission at 1.4GHz originates via synchrotron emission,
either in supernovae remnants \citep[e.g.][]{Condon90,Condon92,Weiler02} --
which correlates with star formation -- or in interactions between jets and the
ambient medium
\citep[e.g.][]{Burbidge56,LNP_relativisticphysics_review_02,Kaiser06}. We are
primarily concerned with early-type galaxies, most of which are not forming
stars. However, to ensure that we are dealing with nuclear radio sources which
are unlikely to be due to star formation, we restricted ourselves to those
galaxies whose radio emission was significantly stronger than what one would
predict from the optically determined star formation rate; we refer to these as
``radio-AGN'' sources. Specifically, we used the fiber-based B04 star-formation
rates (which are insensitive to aperture corrections and well matched to the
average compact nuclear radio source), and then estimated the expected 
star-formation-based radio luminosity using the relation of \citet{Hopkins01}:
\begin{equation}
  L_{\rm 1.4GHz,SF} \; = \; 1.26 \times 10^{21} \times {\rm SFR} (\Msolyr) \, {\rm W \, Hz^{-1}},
\label{equ:radio-agn}
\end{equation}
converted to a \citet{Kroupa01} IMF for consistency with B04. Since Hopkins et
al.\ found approximately an order of magnitude scatter in their relation, we
impose a conservative limit of $L_{\rm 1.4GHz,FIRST} \geq 10 \times L_{\rm 1.4GHz,SF}$
in order to identify bona-fide radio-AGN sources. As an example, a source at the
detection limit of the FIRST survey (flux of 1 mJy) at the high-redshift end of
our sample ($z \sim 0.05$) could be explained by a central SFR of $\sim 7.5
\Msolyr$, but would only be counted as a radio-AGN if the measured SFR was  $<
0.75 \Msolyr$.

The resulting 140 galaxies in the SDSS-RC3 sample for which we found
radio-AGN sources are indicated in Figure~\ref{figure:bpt} by the black dots inside
the galaxy symbols. Most elliptical and S0 galaxies with radio-AGN sources
in our sample are LINERs, which suggests an active SMBH is indeed present.

\begin{figure*} 
\epsscale{1.00}
\plotone{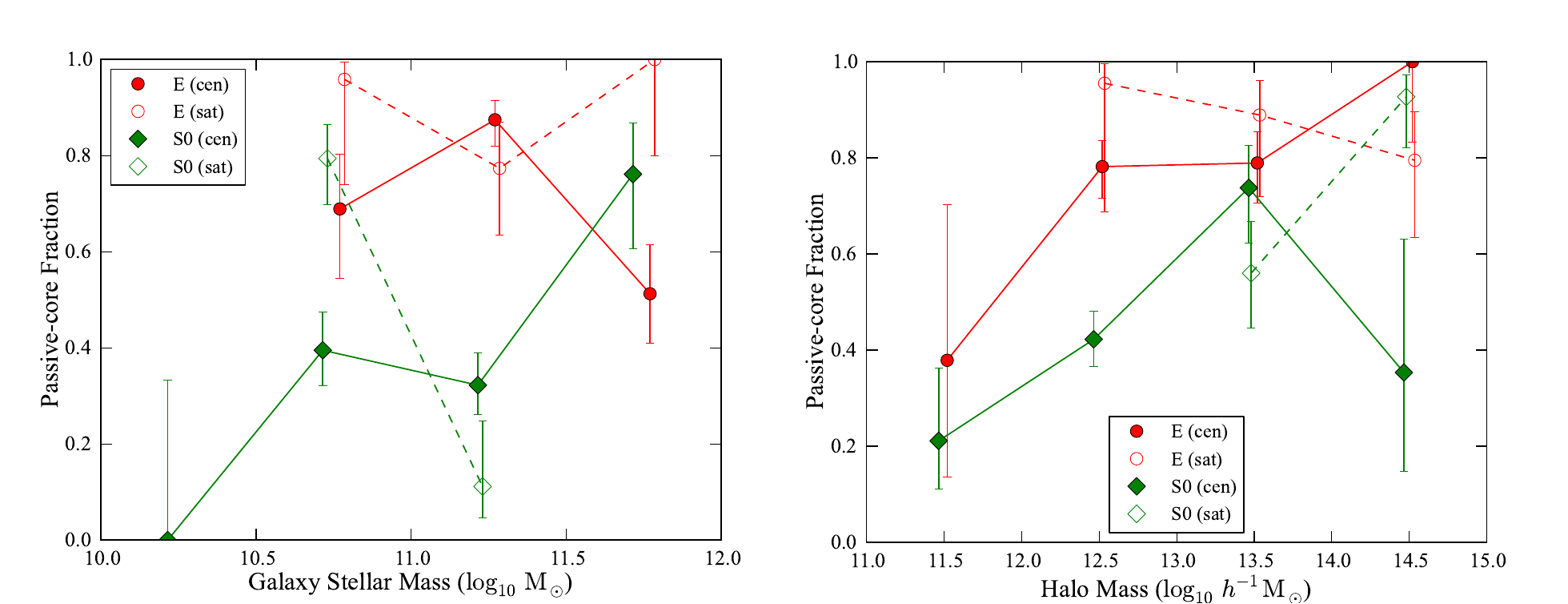} 
\caption{Fractions of $\MB < -19$ elliptical (red circles) and S0
(green diamonds) galaxies with passive cores, separately for central
galaxies (filled symbols) and satellite galaxies (open symbols), using
selection and $V/V_{\rm max}$ weights. This is presented in bins of stellar
mass (left panel) and halo mass (right panel). Ellipticals, and low mass
satellite S0s of high-mass halos, have predominantly passive cores. }
\label{figure:fpES0msmh} 
\end{figure*}

\begin{figure*} 
\epsscale{1.00}
\plotone{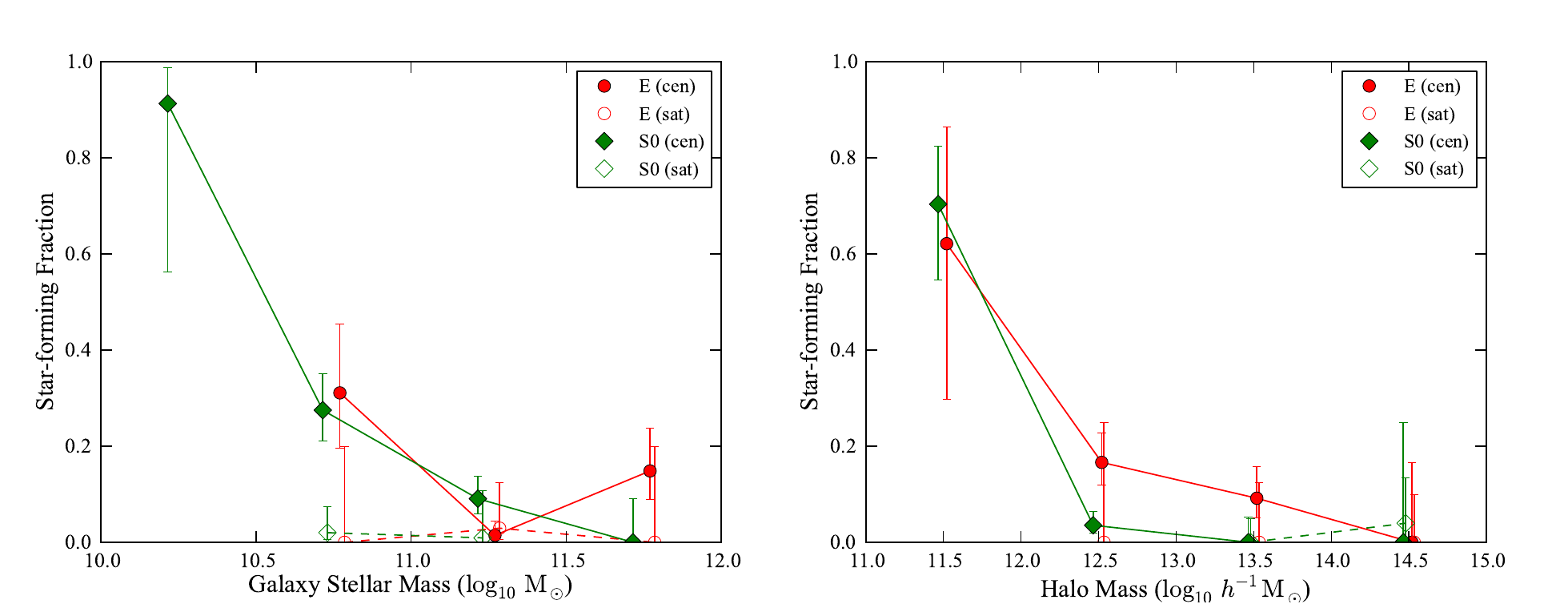}
\caption{As Figure~\ref{figure:fpES0msmh}, but for star-forming
galaxies. Only the central S0s of low mass halos (and few ellipticals or
satellite S0s) have core spectra which put them in the star forming
category.} \label{figure:fsfES0msmh} 
\end{figure*}

\begin{figure*} 
\epsscale{1.00}
\plotone{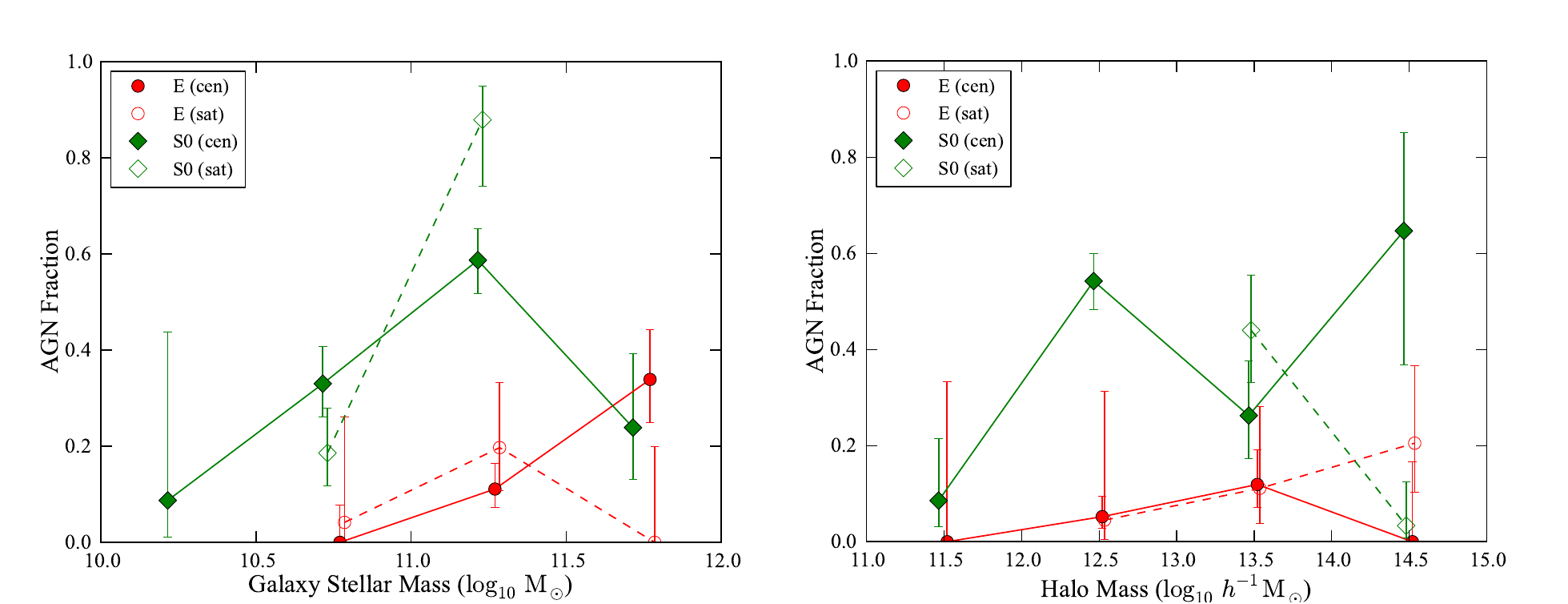}
\caption{As Figure~\ref{figure:fpES0msmh}, but for emission-line AGN.
Early-type AGN (typically LINERs) live mainly in S0s rather than
ellipticals, and in high-mass galaxies. There is no obvious dependence
on environment ($\Mhalo$ or central/satellite status).}
\label{figure:fagnES0msmh} 
\end{figure*}

\begin{figure*} 
\epsscale{1.00}
\plotone{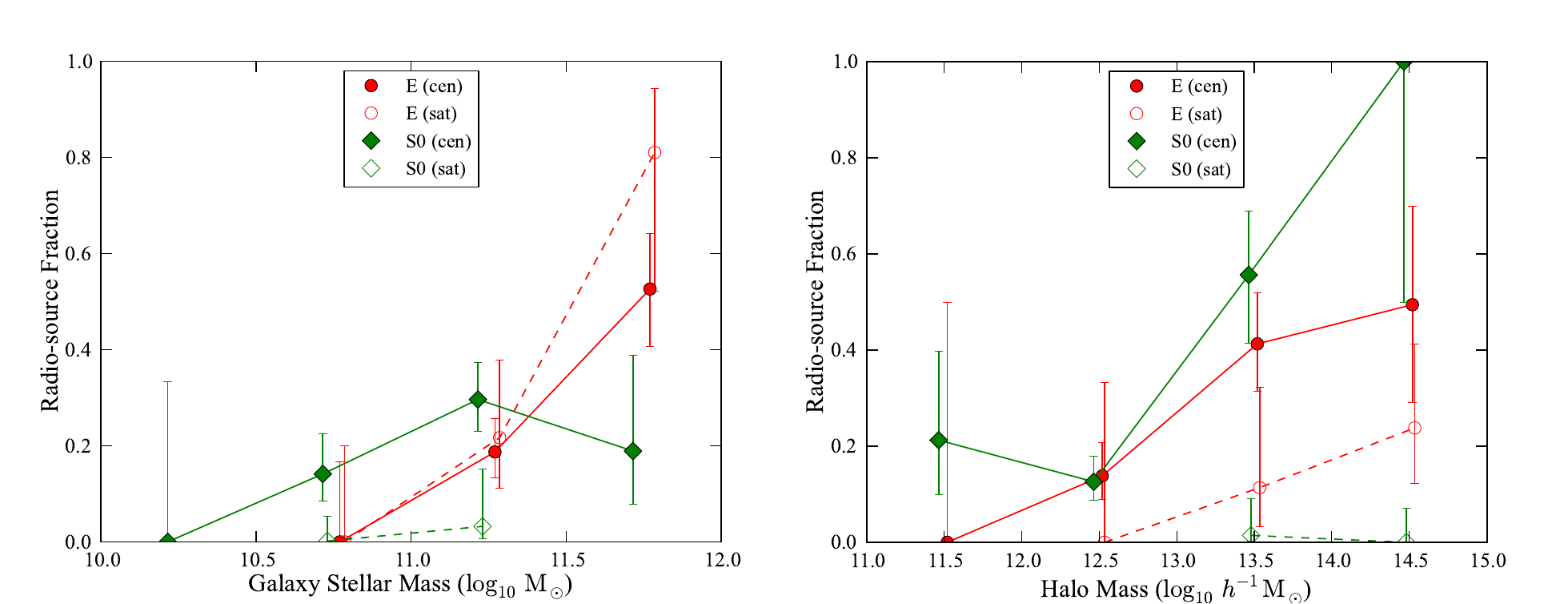}
\caption{As Figure~\ref{figure:fpES0msmh}, but for radio-AGN sources
(nuclear radio sources too bright to be explained by star formation). These
  live primarily in the central, massive galaxies of massive halos
  (elliptical and S0) or in massive satellite ellipticals.}
\label{figure:fradioES0msmh}
\end{figure*}

Figures~\ref{figure:fpES0msmh} to \ref{figure:fradioES0msmh} show the fractions
of elliptical (red circles) and S0 (green diamonds) galaxies which belong to
each spectroscopic class (passive-core, star-forming, AGN), or with radio-AGN
sources, divided into central (solid symbol) and satellite (open symbol)
categories. This fraction is plotted both against stellar mass $\Mstellar$ (left
panels) and halo mass $\Mhalo$ (right panels). These figures illustrate how the
nuclear spectroscopic and radio properties of galaxies depend upon their
morphology, mass and environment. In contrast to the plots in
Section~\ref{sec:lumdep} and \ref{sec:morphcensat}, we extend the stellar-mass
plots here down to $10^{10} \Msol$, because the bias towards spirals at low
stellar masses (Section~\ref{sec:masslimits} and Figure~\ref{figure:mbmstellar})
is no longer relevant.

Figure~\ref{figure:fpES0msmh} shows that the cores of elliptical
galaxies (both central and satellite) are much more freqently passive
than S0s ($70^{+5}_{-6}\%$ for central Es compared to $34 \pm 5$\% for
central S0s, $90^{+5}_{-9}\%$ for satellite Es compared to
$65^{+8}_{-9}\%$ for satellite S0s). The fraction of passive-core S0s is
highest for low mass satellites, which all live in massive $\Mhalo >
10^{13} \, h^{-1} \Msol$ halos (section~\ref{sec:morphcensat}).

Figure~\ref{figure:fsfES0msmh} shows that the only significant population of
early-type galaxies with core spectra indicating star formation are central S0s
with low stellar masses in low-mass halos ($\beta = -9 \pm 4.1, \Plog = 0.029$
versus $\Mstellar$ and $\beta = -8.6 \pm 4.1, \Plog = 0.04$ versus $\Mhalo$). 
Selection effects are relevant, such that low mass star-forming S0s only make it
into our {\it B}-selected sample because they are bright in that band. However,
this population extends into bins of {\it stellar mass} above our threshold
$\Mstellar > 10^{10.5} \Msol$, suggesting a physical truncation of star
formation (and/or the presence of harder, AGN-like ionization) in central S0
galaxies of halos $\Mhalo > 10^{12} \, h^{-1} \Msol$. There are no notable
trends for star-forming cores in elliptical galaxies.

Figure~\ref{figure:fagnES0msmh} shows that emission-line AGN (mostly LINERs,
Figure~\ref{figure:bpt}) are found much more frequently in S0s than in
ellipticals (combined central plus satellite populations: $9 \pm 3$\% for
ellipticals and $34 \pm 4$\% for S0s), and their frequency increases with
stellar mass (notably for S0s, although this is not formally significant: $\beta
= 2.8 \pm 1.5, \Plog = 0.067$). However there is no measurable dependence on
environment for AGN fraction, either in terms of halo mass or central versus
satellite status.

Figure~\ref{figure:fradioES0msmh} further explores the AGN theme in terms of
radio emission. The fraction of galaxies with radio-AGN sources increases with
both galaxy mass and halo mass. For central ellipticals, we find {$\beta = 2.8
\pm 1.4, \Plog = 0.052$ for radio-AGN fraction versus galaxy mass and $\beta =
1.5 \pm 0.53, \Plog = 0.0074$ versus halo mass. The right-hand panel of the
figure shows that for a given halo mass, central galaxies are more likely to
host radio-AGN sources than satellite galaxies; the least likely hosts are
\textit{satellite} S0s.

\subsection{Dependence on Environment and Hubble Type of Bar and Ring
Fractions}

Roughly two-thirds of local spiral galaxies -- and a smaller fraction of
S0 galaxies -- are barred
\citep[e.g.,][]{Eskridge00,Menendez-Delmestre07}. Although $N$-body
simulations have long shown that bars can form spontaneously in isolated
disks, simulations have also shown that bar formation can be triggered
by tidal interactions
\citep[e.g.,][]{Noguchi87,Salo91,Noguchi96,Berentzen04}. It is therefore
plausible that local environment might influence the frequency (and
possibly the size or strength) of bars in disk galaxies. Similarly,
although outer rings are well understood as being primarily due to the
interaction of a bar's Outer Lindblad Resonance with gas in the disk
\citep[e.g.,][]{Buta96}, the fact that they are features of the outer
disk means they are in principle more vulnerable to interactions than
other, more central structures. Thus, we might expect that local
environment could also influence the frequency of outer rings.

To estimate the bar fraction, we consider both strong (RC3 class SB) and
weak (RC3 class SAB) bars in disk galaxies (S0s and spirals considered
separately). We also restrict the sample to relatively face-on galaxies:
those with RC3 axis ratios $a/b \leq 2.0$ ($r25 \leq 0.301$).  The
latter restriction excludes highly inclined galaxies, where optical bar
detection becomes difficult or impossible. We warn the reader in advance
that we are probably \textit{underestimating} the true bar fraction,
since some bars will have been missed due to dust obscuration
\citep[see, e.g.,][]{Eskridge00} and weaker and smaller bars will have
been difficult to identify due to resolution effects. Our analysis
should thus be seen as investigating the possible effects of group
environment on large, strong bars, rather than on all possible bars.

We do find some evidence for trends in bar fraction with galaxy mass.
Figure~\ref{figure:fbar} suggest that the spiral bar fraction increases with
galaxy stellar mass, although the significance of this depends on which set of
stellar masses we use (e.g., $\Plog = 0.020$ using G05 masses, but only 0.058
with our preferred Z09-based masses, and 0.047 for the Y07 masses).  A similar
trend trend may exist for central S0s ($\Plog = 0.017$ using Z09-based masses,
with similar values for the other mass estimates). Such a trend would be
consistent with the findings of \citet{Nair10bars} for a similar mass range.

We find \textit{no} evidence for trends in bar fraction versus halo
properties, either as a whole or when considering central and satellite
galaxies separately (Figure~\ref{figure:fbar}, right panel). This is
consistent with other recent studies of bar fraction with environment.
For example, \citet{Marinova09} found no difference in bar fraction
(determined using optical HST images) across a range of local
environments in the Abell 901/2 Supercluster, and \citet{Aguerri09}
found no evidence for a dependence of bar fraction (determined from SDSS
images) with local environment; both of these studies used local
projected surface density of galaxies as the ``environment''.
\citet{Li09} also found no difference in the clustering properties of
barred versus unbarred galaxies.

For outer rings and pseudorings, we restrict ourselves to S0--Sbc galaxies,
which is where almost all such rings are found \citep[][]{Buta96}. We also
consider rings and pseudorings separately, in part because previous work by
\citet{Elmegreen92} suggested there might be divergent trends for the two
subtypes. There is no clear evidence for any trend of outer ring or pseudoring
frequency with halo mass. The right-hand panel of Figure~\ref{figure:fouterring}
appears to suggest a decreasing frequency with higher halo mass, at least for
central galaxies, but this is not statistically significant, even if we lump
outer rings and pseudorings together ($\Plog = 0.97$ for central galaxies,
$\Plog = 0.25$ for all galaxies). What \textit{does} seem to be present is a
decrease in outer ring or pseudoring frequency with increasing galaxy stellar
mass, at least for central galaxies (left panel of
Figure~\ref{figure:fouterring}).  These trends are relatively shallow, but
statistically significant (e.g., $\Plog = 0.00011$ for central galaxies, lumping
both outer rings and pseudorings together). Since outer rings are usually
associated with bars \citep[e.g.,][]{Buta96}, and since the frequency of bars
apparently {\it increases} with stellar mass as noted above, the lack of outer
rings in more massive galaxies is highly significant.

The (tentative) absence of environmental trends for outer rings appears
to contradict what \citet{Elmegreen92} found: they argued that outer
rings decreased in frequency for denser environments, while the
frequency of pseudorings increased. One difference is that their main
analysis was restricted to strongly barred S0 + S0/a galaxies only, and
it is not clear how to compare their ``field'', ``pseudo-field''
(possible group members), ``group'', and ``binary'' classifications with
our group halo properties. For example, a ``binary galaxy'' system could
be low-halo-mass group with two significant members, or a subset of a
larger group with a higher halo mass. It is also worth noting that the
high outer-ring fraction reported by Elmegreen et al.\ for field
galaxies is based on very small sample sizes (3 field galaxies, 4
pseudo-field galaxies).


\begin{figure*} 
\epsscale{1.00} 
\plotone{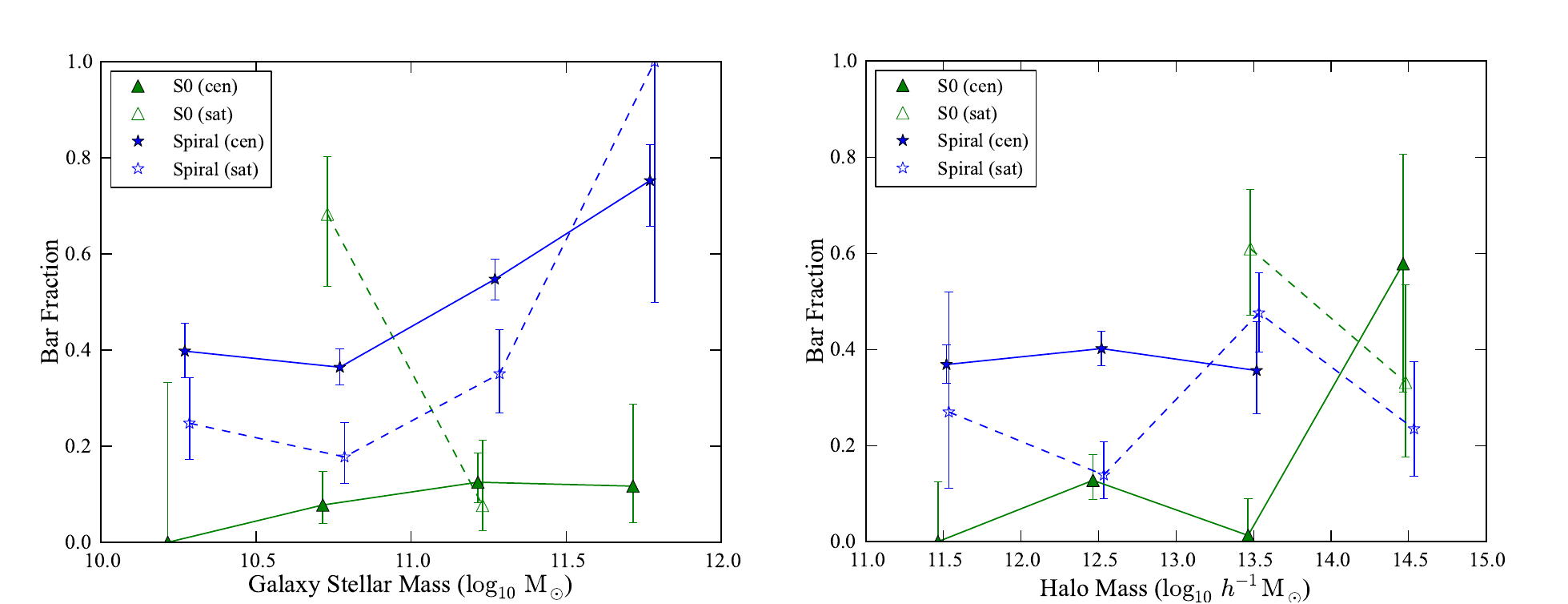} 
\caption{Fractions of strong + weak bars in ($\MB < -19$) S0 (green
triangles) and spiral galaxies (blue stars) as a function of the
galaxies' stellar and halo masses, with selection and $V/V_{\rm max}$
weights applied. Filled symbols indicate central galaxies and hollow
symbols indicate satellite galaxies. In this plot, we consider only
galaxies with isophotal axis ratios $a/b < 2$ (i.e., relatively
face-on). } \label{figure:fbar}
\end{figure*}

\begin{figure*} 
\epsscale{1.00}
\plotone{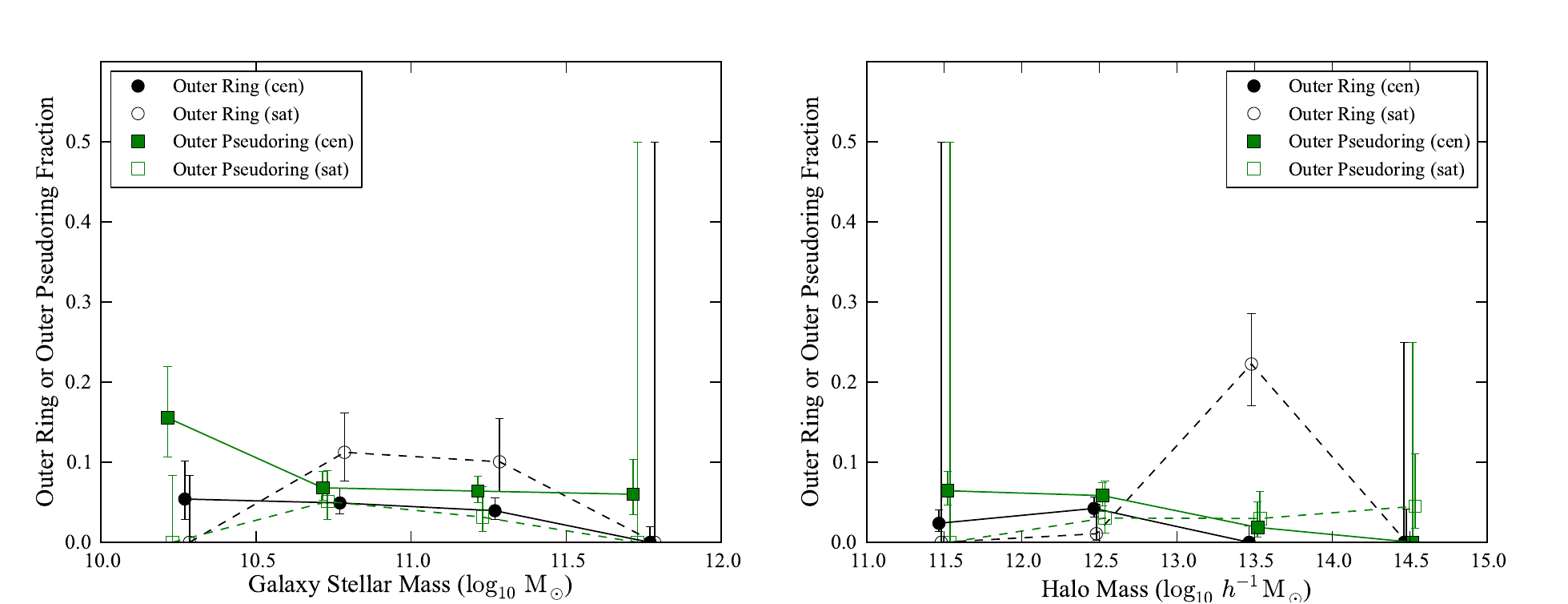} 
\caption{As for Figure~\ref{figure:fbar}, but now showing
fractions of S0--Sbc galaxies which contain either outer rings (black
circles) or outer \textit{pseudo}rings (green squares) as a function of
the galaxies' stellar and halo masses. Filled symbols indicate central
galaxies and hollow symbols indicate satellite galaxies.}
\label{figure:fouterring} 
\end{figure*}

\begin{figure*} 
\epsscale{1.00}
\plotone{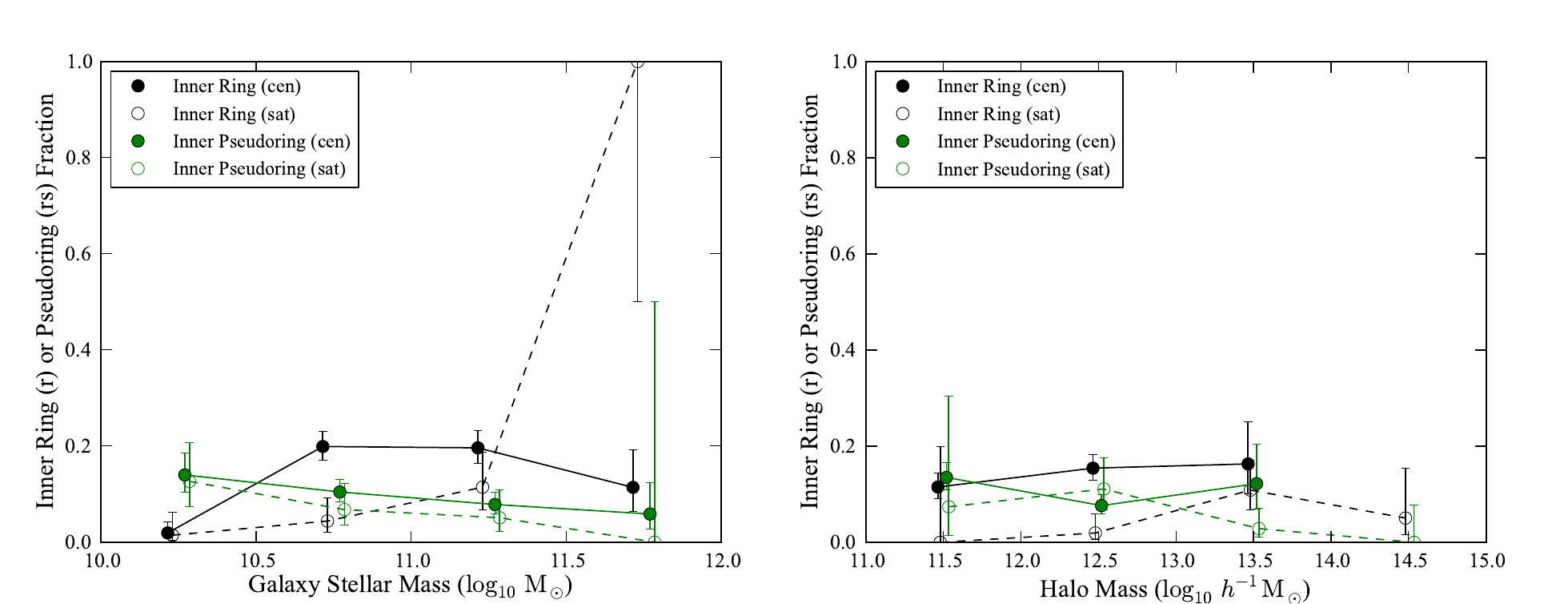} 
\caption{As for Figure~\ref{figure:fbar}, but now
showing fractions of spiral galaxies which contain inner rings (RC3
``r'' classification, black circles) or inner pseudorings (``rs'', green
circles) as a function of the galaxies' stellar and halo masses. In this
plot, we consider only galaxies with isophotal axis ratios $a/b < 2$
(i.e., relatively face-on). Filled symbols indicate central galaxies and
hollow symbols indicate satellite galaxies.} \label{figure:finnerring}
\end{figure*}

\begin{figure*} 
\epsscale{1.00}
\plotone{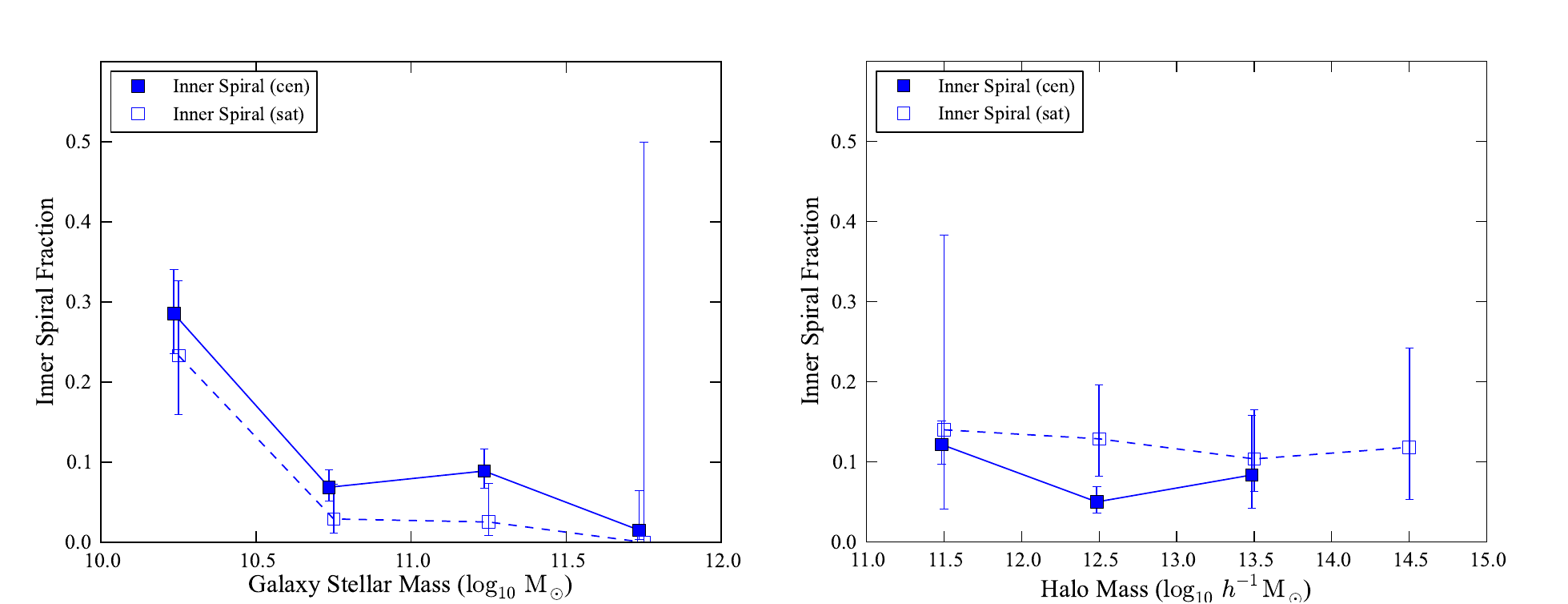} 
\caption{As for Figure~\ref{figure:finnerring}, but
now showing fractions of galaxies with inner \textit{spirals} (RC3 ``s''
classification).} \label{figure:finnerspiral} 
\end{figure*}

The RC3 catalog also provides classifications for \textit{inner}
ring/spiral structure in spirals, specifying whether the main disk
(outside the bar, if present) is purely spiral (s), contains an inner
ring (r), or has an intermediate broken-ring or pseudoring appearance
(rs); see, e.g., Figure~1 of \citet{Buta94}.  We have looked for
possible correlations of r/rs/s frequency with environment in our
sample, but find no evidence for any clear trends with halo mass,
dispersion, or number of group members (Figures~\ref{figure:finnerring}
and \ref{figure:finnerspiral}). We note that the lack of any trend for the
inner-ring fraction is possibly in conflict with \citet{Madore80}, who
found that galaxies with inner rings had fewer close companions
(galaxies at projected distances $< 50$ kpc) than inner-spiral or
inner-pseudoring galaxies. However, Madore pointed out that the number
of close companions, so defined, seemed independent of whether or not a
given galaxy was in a group.

We \textit{do} find that the frequency of inner rings increases with galaxy
luminosity and mass ($\beta = 2.4 \pm 0.36, \Plog = 1.8 \times 10^{-11}$ for
$\Mstellar$, with similar slopes and similarly small values of $\Plog$ for G05
and Y07 stellar masses). Given that inner rings are more common in early-type
spirals \citep{DeV80}, this trend could be a side effect of the tendency of
early type spirals to be more massive than late-type spirals; it could also be
due to the apparent increase in bar fraction with stellar mass noted above,
since inner rings are usually due to bar-related resonances \citep{Buta96}.
Figure~\ref{figure:finnerspiral} does suggest a corresponding decrease in
inner-spiral fraction with galaxy mass, though this is not statistically
significant ($\beta = -0.7 \pm 0.61, \Plog = 0.25$).

\section{Discussion}\label{sec:disc}

\subsection{Where and How Are Elliptical Galaxies Formed?}\label{sec:ellipticals}

Figure~\ref{figure:ftypemstar} shows that the global fraction of
elliptical galaxies increases with stellar mass, while
Figure~\ref{figure:ftype-combined} shows no such trend with halo mass,
velocity dispersion or number of group members.  This could naively be
interpreted as evidence for a purely mass-dependent formation of
ellipticals, independent of environment. However,
Figure~\ref{figure:ftypemhalocensat} shows that a strong trend with halo
mass does exist when only \textit{central} galaxies (the most massive
ones in their group) are considered. No such trend is seen for
satellites; the fraction of ellipticals for satellites correlates only
with stellar mass. The similar stellar mass dependences of central and
satellite ellipticals, and the relatively infrequent occurance of
satellite ellipticals, argues for a common parent population for both
types.

Simulations show that disks are largely destroyed by major mergers, leading to
the formation of elliptical galaxies \citep[e.g.][]{Barnes88}.  These events
preferentially occur at the centers of halos: dynamical friction brings
satellite galaxies to the bottom of the potential well where they merge with the
central galaxy. Cosmological simulations show that mergers between the
subhalos hosting satellite galaxies are rare \citep{Angulo09}.  The highest
mass halos have the richest halo merger history, and the central galaxies of
these halos are the most massive galaxies in the Universe (the most extreme
examples being cD galaxies in clusters), with the most extensive merger
histories of any galaxy \citep[e.g.][]{deLucia07}.

If all ellipticals are formed as (or transformed into) ellipticals while
they are still central galaxies within their own halos, then the
observed correlations are perfectly consistent with our understanding of
how ellipticals form via mergers \citep{deLucia11}. The more massive a
halo, the more likely its central galaxy is to have undergone multiple
major mergers, and thus the more likely it is to be an elliptical; such
galaxies will naturally also tend to be more massive. 
\textit{Satellite} ellipticals are galaxies which formed as ellipticals
at the center of their progenitor halos -- thus partaking in the general
trends just outlined -- and were subsequently accreted \textit{as
ellipticals} onto their current halos. Correlations with stellar mass
will therefore persist for satellite ellipticals, whilst their previous
(central-galaxy) correlation with halo mass is lost.

In this scenario, most properties of elliptical galaxies (in addition to
mass and morphology) are determined by their formation as central galaxies, and
subsequently frozen at the time of their accretion onto larger halos as
satellites. Thus, aside from possible differences in mean stellar age, satellite
and central ellipticals should follow the same general scaling relations.
Studies have found that the well-known elliptical-galaxy scaling relations are
indeed largely independent of environment, These include the slope of the
fundamental plane (e.g., \citealp{DelaRosa01,Reda05,Bernardi06}; but see also
\citealp{DOnofrio08,LaBarbera10}), the color-magnitude and Kormendy relations
\citep[e.g.,][]{Hogg04,Reda05}, and luminosity-size relations
\citep{Nair10sizes}. \citet{Guo09} and \citet{Weinmann09} studied galaxies using
the Y07 group catalog, specifically contrasting mass-matched central and
satellite galaxies, and found no size or structural differences for
``early-type'' galaxies. Note that many of these studies lumped elliptical and
S0 galaxies together, so there is in principle the possibility of confusion if
ellipticals and S0s have trends that happen to cancel out when they are combined.

While there \textit{is} evidence for \textit{age} differences between ellipticals
in different environments \citep[e.g.,][]{Thomas05,Bernardi06,Saglia10}, this is
still consistent with the overall picture. Ellipticals in high density environments
are more likely to be satellite galaxies within massive halos, with properties
frozen at the time of the accretion.


The picture outlined above implicitly assumes that once an elliptical has formed
at the center of its halo, it is able to \textit{remain} an elliptical. In the
case of ellipticals created by ``wet'' (gas-rich) mergers, the problem is how to
prevent significant residual gas from continuing to form stars; quasar-mode
feedback is a promising solution
\citep[e.g.,][]{Granato04,Springel05,Hopkins06}. A more general, long-term
problem is that posed by the presence of hot gas in the halo. Since the center
of a halo is the natural destination for halo gas that is able to cool, some
mechanism must exist for suppressing such cooling -- otherwise, cool gas would
accumulate in the center of the halo and potentially form a new stellar disk.
(Note that \textit{satellite} ellipticals do not suffer from this problem, since
they do not sit at the centers of halos.)

This cool, low entropy gas has been observed in {\it some} massive clusters --
but it is not ubiquitous; clusters with no cooling flows are in the majority
\citep[e.g.][]{Nesci91,Cavagnolo09}.  The existence of cooling gas appears
to be a requirement for central cluster galaxy to host either star
formation \citep[e.g.][]{Donahue10,Hicks10} or radio AGN \citep[e.g.][]{Sun09}.
It has been proposed that energy from a radio jet can offset cooling in a
cluster and that this suppresses the growth of the central galaxy. Cavities
containing low-density, hot gas have been observed spatially coincident to the
radio lobes, and the total energy required to create such cavities is typically
enough to balance cooling
\citep[e.g.][]{McNamara00,Fabian00,Dunn06,Cavagnolo10}.

If radio-mode AGN feedback is also applicable for galaxies in lower-mass halos,
then it could prevent significant cooling onto $\sim L_{*}$ galaxies,
suppressing their growth and allowing models of galaxy formation to match the
high-mass exponential cutoff seen in the galaxy mass function
\citep{Bower06,Croton06}. Cavities \textit{have} been observed in the hot gas
component of galaxy groups \citep{Dong10} and even individual elliptical
galaxies \citep{Baldi09}. The high fraction of radio sources we find for central
ellipticals and S0s in halos with masses of $10^{13}$--$10^{14.5}\, h^{-1}
\Msol$ \citep[Figure~\ref{figure:fradioES0msmh}; see also][]{Best07,Pasquali09}) is
potentially further support for the possibility that radio-mode feedback
operates in halos of these masses. It is also possible that other forms of AGN
feedback can suppress star formation in lower-mass halos; for example,
\citet{Schawinski09} have found evidence that low-luminosity AGN activity is
associated with the disappearance of central molecular gas in S0 and elliptical
galaxies.

More detailed models will be necessary to determine whether the fractions
of central galaxies with elliptical as opposed to disk
morphology, and their dependence on halo mass, can be {\it quantitatively}
explained in the context of the expected merger history and suppression of disk
re-formation (Wilman et al., in prep).

\subsection{Where Are S0s Formed?}

As with ellipticals, the dependence on halo mass of the S0 fraction is
very different for satellite and central galaxies.  The remarkable
change in satellite S0 frequency, which jumps from
$0^{+2.6}_{-0}$\% in lower-mass halos to
$69.0^{+4.3}_{-4.6}$\% for halos with
masses $> 10^{13} h^{-1} \Msol$, suggests that spiral arms are often
suppressed in a disk galaxy once it is accreted onto a halo,
\textit{if} the halo is more massive than $10^{13} \, h^{-1} \Msol$.

In fact, we can argue that the majority of all present-day satellite S0 galaxies
became S0s \textit{after} they were accreted into halos more massive than $\sim
10^{13} \, h^{-1} \Msol$. Since $z = 0$ halos with $\Mhalo < 10^{13} \, h^{-1}
\Msol$ do not have satellite S0s, any pre-existing S0s which formed in less
massive progenitor halos and then fell into massive halos ($\Mhalo > 10^{13} \,
h^{-1} \Msol$) must have originally been \textit{central} galaxies.  Assuming
that these progenitor halos had a distribution of central morphological types
similar to what we see today (if anything, they probably had lower S0 fractions
in the past), then the mean central S0 fraction for the accreting progenitor
halos would be $\sim f_{\rm S0}^{\rm cen} = 20.1\pm1.6$\%.  The fraction of S0s
in massive halos which are {\it post-processed} is therefore:
\begin{equation}
  f_{\rm S0}^{\rm post} \, = \; f_{\rm S0}^{sat} \, - \, f_{\rm S0}^{\rm cen},
\label{equ:S0postprocessed}
\end{equation}
where $f_{\rm S0}^{\rm post}$ is the fraction of massive-halo
satellite galaxies which are post-processed S0s, and $f_{\rm S0}^{\rm
  sat}$ is the fraction of massive-halo satellites which are S0s. The
present-day S0 satellite fraction ($f_{\rm S0}^{\rm sat} \, = \, $
$69 \pm 4$\%) thus requires that
almost three quarters of these
galaxies (i.e., $49 \pm 5$\% of
$\Mhalo>10^{13} \, h^{-1}\Msol$ satellite galaxies, or $71 \pm
  8$\% of satellite S0s) fell into their
present-day halos \textit{as spirals}, becoming S0s during or after
the accretion process.

We can go one step further, and apply
equation~\ref{equ:S0postprocessed} as a function of galaxy stellar
mass. We limit this exercise to the stellar mass range $10^{10.5}\Msol
\leq \Mstellar \leq 10^{11.5}\Msol$ for two reasons: first, as
previously noted, S0s are lost due to selection effects below
$10^{10.5}\Msol$; second, there are no satellite S0s in our sample with
masses $> 10^{11.5} \Msol$ (see Figure~\ref{figure:ftypemhalocensat}).
For simplicity, we assume that there is no significant stellar-mass
change during or after the accretion process. The upper panel of
figure~\ref{figure:postprocessedS0s} shows how $f_{\rm S0}^{\rm cen}$
and $f_{\rm S0}^{\rm sat}$ depend on stellar mass.  We also show the
equivalent fractions for spiral galaxies: i.e., $f_{\rm Sp}^{\rm cen}$
and $f_{\rm Sp}^{\rm sat}$. What is striking about this plot is the
apparent stellar-mass trend. For $\Mstellar < 10^{11} \Msol$, the
satellite S0 fraction is clearly too high to be explained by the
pre-processed S0 population alone, requiring substantial conversion of
accreted spirals into post-processed S0s. At the high-mass end, on the
other hand, most or all of the satellite S0s can be explained as
pre-processed S0s, with little or no conversion of spirals required.
This is consistent with the fact that the satellite \textit{spiral}
fraction in the highest-mass bin is basically the same as the central
spiral fraction, suggesting that $\sim$ all of the highest-mass spirals
have remained spirals after accretion.  This can be compared with
\citet[][esp.\ their Figure~8]{VanDenBosch08}, who apply a similar
argument to suggest that the high fraction of massive, $\Mstellar
\gtrsim 10^{11} \, h^{-2} \, \Msol$ galaxies which have {\it red}
colors is largely due to their pre-processing as central galaxies.

The lower panel of figure~\ref{figure:postprocessedS0s} makes some of this
more explicit. Here we show our estimate for the fraction of {\it pre-processed
S0s}, assumed to be equal to the fraction of central S0s in the top panel. We
also estimate the fraction of {\it post-processed S0s} using
equation~\ref{equ:S0postprocessed}; as hinted in the top panel, the
fraction of satellite S0s which are post-processed increases to lower masses. 
Finally, we also plot the fraction of \textit{``missing''
satellite spirals}, which is the fraction of central spirals minus the fraction
of satellite spirals in $\Mhalo>10^{13} \, h^{-1}\Msol$ halos. This quantity
represents the fraction of accreted spirals which are no longer present
\textit{as spirals}; these are assumed to have transformed into S0s or else
merged with other galaxies. In each mass bin, the fractions of
post-processed S0s and missing spirals are \textit{equal} within the errors,
strongly suggesting that the missing spirals have indeed been converted into
(post-processed) S0s.

\begin{figure*}
  \epsscale{1.00}
  \plotone{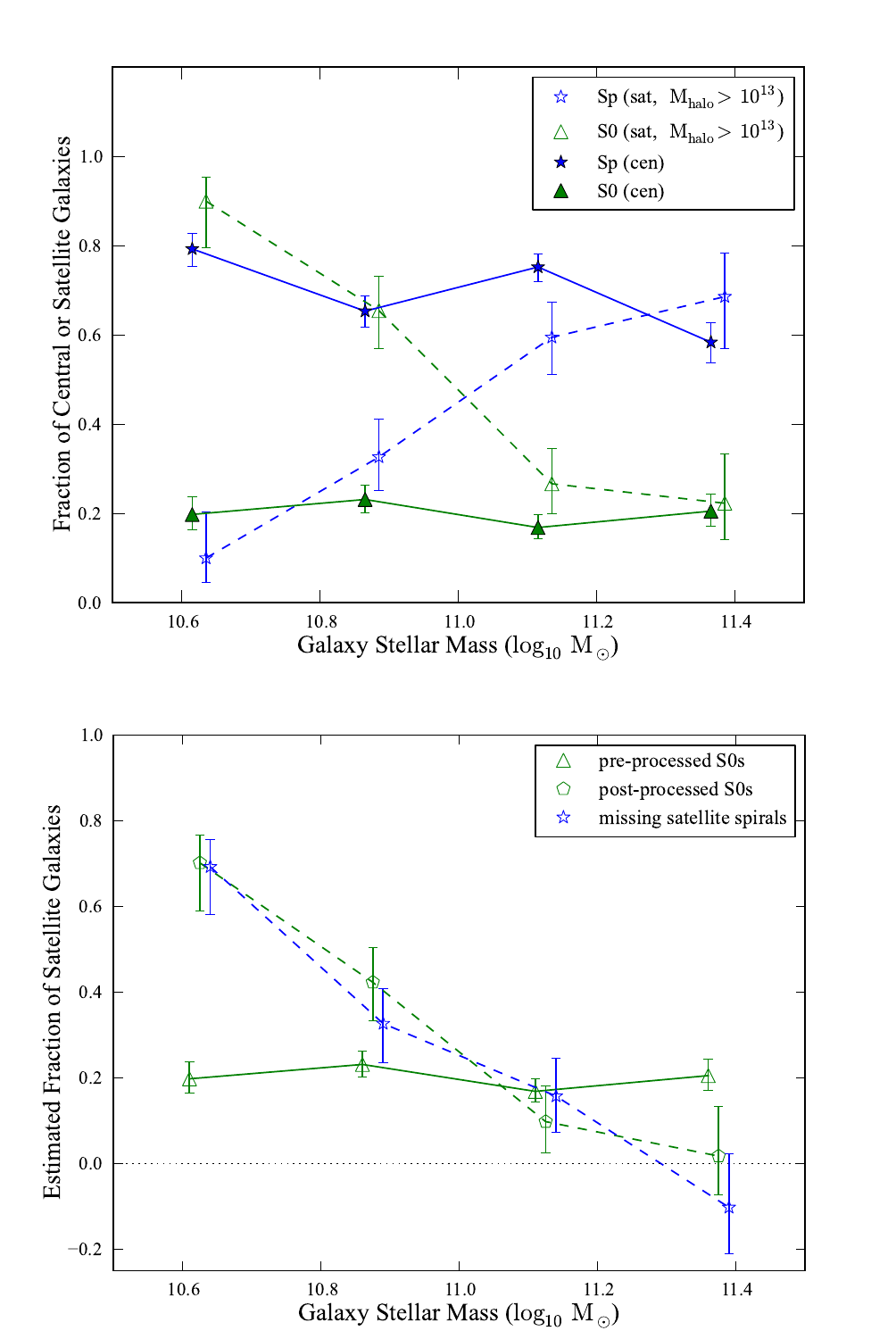}
\caption{{\bf Top:} Observed fractions of S0 and spiral galaxies as a function
of stellar mass $\Mstellar$ for central galaxies (solid points and lines)   and
for the satellites of massive ($\Mhalo > 10^{13} \, h^{-1}\Msol$) halos (open
points and dashed lines). {\bf Bottom:} Estimated fractions for the satellite
population of massive halos, divided into ``pre-processed'' S0 galaxies
(pre-existing S0 galaxies accreted by the halo) and ``post-processed'' S0
galaxies (accreted as spiral galaxies and subsequently transformed). These
fractions assume that the morphological fractions of galaxies at the time of
accretion matches the currently observed \textit{central}-galaxy fractions (top
panel).  We also show the fraction of ``missing satellite spirals''---these are
galaxies accreted as spirals which are no longer present \textit{as spirals}
(i.e., they are presumably either transformed into S0's or merged to form
ellipticals).}
  \label{figure:postprocessedS0s}
\end{figure*}

\subsection{How Are S0s Formed?}

The persistence and visibility of spiral arms clearly correlates with
the presence of gas and star formation: spiral galaxy disks are forming
stars, while S0 disks are typically passive (e.g.,
Figure~\ref{figure:fSFSphtar}). The absence of spiral arms in S0
galaxies can be explained by the combination of increased random motions
of disk stars with age, which erases existing spiral patterns, and lack
of young stars in regular, ``cold'' orbits, which would otherwise
maintain or reform spiral patterns  \citep[see e.g. section~6.1 of][and
references therein]{Sellwood11}.

Star forming galaxies at z $\sim 0$ typically have enough atomic and
molecular gas to maintain star formation for only another $\sim 3$ Gyr
on average at present rates \citep[for the statistics of local galaxies
see][]{Saintonge11}.  Ongoing disk star formation for times of order the
Hubble time thus requires the availability of additional gas, accreted
from the surroundings. This gas is usually assumed to be shock-heated
upon accretion onto $\gtrsim 10^{12} \, h^{-1} \Msol$ halos, resulting
in a ``hot atmosphere'' which subsequently (in the absence of further heating)
cools onto the galaxy \citep{WhiteFrenk91}.

The transformation of spiral galaxies into S0s therefore requires that star
formation is suppressed, both by the exhaustion or removal of existing disk gas
and by the prevention of further gas accretion from the environment.  Our
results suggest a difference between the transformation of post-processed
satellite galaxies in $\Mhalo > 10^{13} \, h^{-1} \Msol$ halos on the one hand,
and that of \textit{central} galaxies in halos down to our limiting halo mass of
$\sim 10^{12} \, h^{-1} \Msol$ -- \textit{independent} of halo mass -- on the
other hand. We therefore expect different mechanisms to be responsible for S0
formation in these two regimes.

Finally, whatever mechanisms are operating, they also need to explain the
observed {\it structural} differences between spirals and S0s. S0s have
traditionally been characterized as having extremely high $B/T$ ratios
\citep[e.g.,][]{Dressler80}; for example, the compilation of \citet{Simien86}
has a mean $B/T \sim 0.57$ for S0s.  More recent studies which account for
variable bulge profiles (i.e., S\'ersic $r^{1/n}$ instead of de Vaucouleurs
$r^{1/4}$ profiles) and the effects of additional components such as bars have
resulted in lower $B/T$ values for S0s -- but the large study using 2D
decompositions of disk galaxies by \citet{Laurikainen10} still finds that the
mean (and maximum) $B/T$ values for S0s are higher than those of spirals (though
there is considerable overlap, and some S0s have $B/T \lesssim 0.1$); see
Figure~4 of that paper. \citet{Christlein04} constructed separate luminosity
functions for disks and for bulges as a function of $B/T$.  For intermediate,
$0.2 < B/T < 0.7$ galaxies (assumed to be typical for S0s), they found that
while the characteristic luminosity of bulges $L^*_{bulge}$ strongly increases
with $B/T$, the characteristic luminosity of disks $L^*_{disk}$ is almost
constant. Simple toy models are then used to show that this is inconsistent with
changes in $B/T$ resulting from pure disk fading -- but that a model in which
$B/T$ evolves through the growth of bulges is consistent with data. We also note
that \citet{Burstein05} presented evidence indicating the $K$-band luminosities
of local S0s were, on the whole, too high for all of them to be explained by
fading of gas-stripped spirals.

\subsubsection{Satellite S0s}

We have presented evidence that the majority of  S0s in $\Mhalo > 10^{13} \,
h^{-1} \Msol$ halos were likely accreted as spiral galaxies, and have since been
post-processed, leading to their current S0 morphology.  We have also shown 
that lower mass S0s  ($\Mstellar\lesssim 10^{11}\Msol$) are more likely to have
experienced post-processing, whilst the fraction of higher mass S0s is more
consistent with  an accreted field population (pre-processed into S0s as central
galaxies). {Thus, post-processing is more relevant for lower mass disk galaxies,
while higher-mass accreted spirals are more likely to persist as spirals in the
group environment.  This puts potential limits on the post-processing
mechanism(s); in this section, we focus on those which remove gas from the
galaxy, since that is an essential prerequisite for preventing further star
formation and the persistence of spiral structure in the disk.

A promising way of removing gas from spiral galaxies in massive groups and
clusters is to remove it via interactions with the gas of the hot intra-group or
intra-cluster medium (IGM/ICM).  The physical nature of this interaction can --
very broadly -- take three forms: ram-pressure stripping \citep{GunnGott72},
viscous stripping \citep{Nulsen82}, and thermal evaporation \citep{Cowie77}.
Ram-pressure refers to the pressure exerted due to the galaxy's motion relative
to the IGM/ICM and acts very quickly in regions of very dense, hot gas such as
the cores of massive clusters \citep[see e.g.][]{Moran07}; ram-pressure
stripping can potentially be strong enough to remove tightly-bound cold gas in
the galaxy disk.  Viscous stripping refers to the slower removal of low density
gas via turbulence at the interface between the galaxy and IGM/ICM. Both of
these mechanisms depend primarily on the IGM/ICM density and the galaxy
velocity.  Evaporation instead removes gas via thermally induced collisions, and
depends on the temperature of the IGM/ICM; it can also act on central galaxies,
and so is not exclusive to satellite galaxies.

There is direct observational evidence for ram-pressure stripping of cold gas
from disk galaxies in the Virgo Cluster \citep[see e.g.][and references
therein]{Gavazzi08,Chung09} and also in the Coma Cluster
\citep[e.g.,][]{Vollmer01}. However, Virgo and Coma are relatively massive
systems (clusters with $\Mhalo \gg 10^{14} \, h^{-1} \Msol$), supporting dense
ICMs and rapid galaxy motions, both of which lead to stronger galaxy-ICM
interactions. Even here, stripping of disk gas in most cases seems to be
partial, acting only on the outer parts of the disk and leaving a more tightly
bound core gas component, which can continue forming stars.  This is also seen
in simulations \citep[e.g.][]{Kapferer09}. The main problem is that it is not
clear whether stripping of tightly bound cold gas can ever operate effectively
in lower-mass clusters and groups (e.g., $\Mhalo \sim 10^{13}$--$10^{14} \,
h^{-1} \Msol$). \citet{Moran07} studied passive spirals and S0s in two clusters
at $z \sim 0.5$ and concluded that ram-pressure stripping was significant only
in the more massive cluster, with its much denser ICM.

A more widely applicable process is ``strangulation'' \citep{LTC}, which refers
to the ram-pressure removal of just the hot gas halo of a galaxy.  Since the
halo gas is much thinner, hotter, and much less strongly bound than the cold
disk gas, it is vulnerable to removal by lower velocities and lower ICM/IGM
densities, making it a plausible mechanism for lower-mass clusters and groups.
Simulations of strangulation find that, as for ram-pressure stripping of the
cold gas, the central, most strongly bound hot gas component can survive the
stripping process \citep{McCarthy08,Kawata08,Bekki09}. This is not necessarily a
problem, however. Partial stripping of the hot gas causes star formation to be
suppressed over longer timescales than if all of the hot gas and most or all of
the cold gas were stripped (e.g., in the standard ram-pressure-stripping
scenario). With little or no cooling of hot gas onto a satellite galaxy, the
remaining cold disk gas will be exhausted in $\sim 3$ Gyr.

Alternatively, a {\it gravitational} mechanism might be responsible for the
post-processing of group galaxies.  Galaxies with an extended history of minor
mergers are likely to form a high $B/T$ remnant with S0 morphology
\citep[e.g.][]{Bournaud07}. However, if galaxies follow dark matter, the
probability of satellite-satellite mergers is low \citep{Angulo09} and so we
expect this to be more important in the case of central S0s (see below). A more
promising, slow-acting mechanism is the cumulative effect of low-velocity tidal
interactions with the other group galaxies \citep{Bekki11}. This is distinct
from the frequent, high-velocity encounters in more massive clusters, often
referred to as ``harassment'' \citep[e.g.][]{Moore98,Gnedin03}. The simulations
of \citet{Bekki11} show that repeated interactions can drive tidal stripping and
compression of the gas, leading to its more rapid exhaustion than in an isolated
case; this process also produces bulge growth via central
star formation -- albeit by only $\Delta (B/T) \lesssim 10$\%. Tidal
interactions also enhance the random motions of stars, which thickens the disk
and helps suppress spiral arms. This mechanism acts preferentially on lower mass
galaxies (which are more easily perturbed) and on galaxies in lower mass groups
($\Mhalo \sim 2 \times 10^{13} \Msol$). Encouragingly, this is consistent with
our population of post-processed S0s (predominantly $\Mstellar < 10^{11}\Msol$
lower panel of Figure~\ref{figure:postprocessedS0s}).

\subsubsection{Central S0s}

In most cases, the \textit{central} galaxy of a halo lives at the bottom of
the global potential well, and has little or no velocity offset with respect to
the hot gas \citep[see][for more detail and caveats]{Skibba10}. Therefore,
stripping cannot operate under these conditions, and we must look for
other mechanisms to suppress star formation in central S0s.

The centers of halos \textit{are} where galaxy mergers predominate. Since
mergers where one or more galaxy is gas-rich can induce starbursts that rapidly
consume the gas (plus potential quasar-mode feedback), this is a potential
route for forming S0s. Minor mergers, in particular, should be more common than
major mergers and have the additional utility of tending to add mass to the
bulge without completely destroying the disk
\citep{Bekki98,ElicheMoral06,Bournaud05,Bournaud07}; this would help
increase the $B/T$ ratio of central S0s. Of course, \textit{major} mergers --- and
multiple minor mergers \citep[e.g.,][]{Bournaud07} --- are more likely to
produce an elliptical remnant. Since higher-mass halos, with their richer merger
history, are probably more likely to have had major mergers for their central
galaxies \citep[e.g.,][]{Wang08}, we would expect the central galaxies of
higher-mass halos to be ellipticals more often, which is indeed the trend we see
for our sample (Figure~\ref{figure:ftypemhalocensat}).

This suggests that the existence of S0 galaxies with $K$-band luminosities
brighter than any spiral galaxy, as pointed out by \citet{Burstein05}, does
\textit{not} mean S0s \textit{cannot} form through disk fading of spirals.
Instead, it suggests that the most massive S0s are a mixture of S0s which are
still the central galaxies within their groups and formerly central
(pre-processed) S0s which were accreted into massive groups and clusters. Less
massive S0s are then more likely to be post-processed spirals.

We are still left with the problem --- similar to that we faced with explaining
the lack of recent star-formation activity in central ellipticals --- of how to
prevent cooling halo gas from accreting onto central S0s and triggering new star
formation. In principle, the same AGN-feedback mechanisms invoked for central
ellipticals (see Section~\ref{sec:ellipticals}) could apply.  We do in fact
find relatively high radio-AGN frequencies in central S0's, especially in more
massive halos (Figure~\ref{figure:fradioES0msmh}). A possible problem is the
lower masses of central black holes in S0s, which could make feedback less
efficient. (Since SMBH mass scales with \textit{bulge} mass rather than total
galaxy mass --- e.g., \nocite{Kormendy01,Kormendy11}Kormendy \& Gebhardt 2001,
Kormendy et al.\ 2011 --- S0s will tend to have smaller SMBHs than ellipticals
of the same stellar mass.)  However, recent simulations by \citet{Gaspari11}
suggest that relatively weak feedback may be all that is needed to keep cooling
flows from developing in groups.    We also see high fractions of optical AGN
in central S0s as well (Figure~\ref{figure:fagnES0msmh}), which could be helpful
given the evidence that low-luminosity AGN may be associated with the
disappearance of central molecular gas even in galaxies having stellar masses of
a few $\times 10^{10} \Msol$ \citep{Schawinski09}.

\subsection{Evolution of the S0 Fraction}\label{sec:whens0formed}







Comparisons with samples at different redshifts require care -- especially where
different selection limits, environmental definitions, and classification
methods are employed. The top panel of Figure~\ref{figure:postprocessedS0s}
indicates that the S0 fraction is sensitive to the mass or luminosity threshold
imposed -- a problem that is accentuated when comparing samples from different
redshifts and with different photometry.

Since most high-redshift studies where the S0 fractions are computed are for
massive clusters, and thus not good direct comparisons with our sample, we
compare our local S0 fractions with the $z \sim 0.4$ group and field sample
presented by \citet{Wilman09}.  In Figure~5 of that paper, morphological
fractions are computed with a luminosity cut of $M_V = -20.53$, for comparison
with cluster fractions from the literature.\footnote{Selection and systematics
are discussed in Section~4.2 of that paper.}. Assuming passive evolution, the
luminosities of ellipticals and S0s will decrease between $z \sim 0.4$ and now;
thus, we need to adjust our local luminosity cutoff accordingly. Based on the
updated calculations of \citet{VanDokkum01}\footnote{Available at
http://www.astro.yale.edu/dokkum/evocalc/}, we estimate $\sim 0.43$ mag fading
in $V$ for E and S0 galaxies, and therefore adopt a local luminosity limit of
$M_{V} = -20.1$ for comparison purposes.\footnote{We compute $V \; = \; g \, -
\,  0.2906 (u \, - \, g) \, + \, 0.0885$ plus the correction for oversubtracted
background -- see Section~\ref{sec:selection}.} Note that star-forming
\textit{spiral} galaxies are not expected to fade as much as elliptical or S0
galaxies; some may even \textit{increase} in luminosity. Consequently, the local
sample may have an \textit{excess} of spirals relative to the higher-redshift
sample, since some spirals with $M_{V} > -20.53$ at $z \sim 0.4$ now have $M_{V}
< -20.1$.

The \citet{Wilman09} statistics are based on two overall classifications of
environment: ``groups'' and ``field''. Their groups do not have halo mass
estimates, but they \textit{do} have velocity dispersions. Since almost all the
Wilman et al.\  groups have dispersion $\gtrsim 175 \kms$, we divide our local
sample into equivalent subsets, with ``groups'' defined as halos having
dispersions $> 175 \kms$ (this amounts to 90 groups containing a total of 185
classified galaxies) and the ``field'' defined as all other halos.

The S0 fraction for the field shows \textit{no} evolution over this
redshift range: $10^{+6}_{-3}$\% at $z \sim 0.4$ versus $11.1^{+1.2}_{-1.1}$\%
at $z \sim 0.02$. But in groups we \textit{do} see some evidence for evolution:
the fraction rises from $28\pm4$\% at $z \sim 0.4$ to $52.3^{+3.9}_{-4.0}$\%
locally. This parallels at least qualitatively the increase in S0 fraction
observed for clusters over the same redshift range
\citep[e.g.,][]{Dressler97,Fasano00,Poggianti09}.  Quantitatively, it also seems
to agree with the evolution in S0 fraction observed by \citet{Just10} for the
lower-mass clusters in their sample (those with velocity dispersion $< 750
\kms$).

\subsection{Differences in AGN/LINER Fraction Between Ellipticals and
S0s}

Figure~\ref{figure:fagnES0msmh} shows that the fraction of emission-line
AGN (mostly LINERS) is similar for central and satellite galaxies, and
does not depend significantly on halo mass. It \textit{does} increase
with stellar mass, and is much higher in S0s than in elliptical galaxies
of the same mass (except at the very highest-mass end).

The mass of central supermassive black holes is known to increase with galaxy
(more properly, bulge) mass \citep[e.g.,][]{Magorrian98,Marconi03,Haring04}. The
increasing fraction of emission-line AGN with galaxy mass could thus potentially
result from more powerful ionizing sources due to higher black hole masses.
However, the mass of an elliptical galaxy is larger than the bulge mass of an S0
of equivalent stellar mass. The implication of a tight bulge mass--black hole
mass relation is that an elliptical galaxy should also have a more massive black
hole. In fact, however, we generally see a higher fraction of both optical AGN
and radio-AGN in \textit{S0} galaxies, (Figures~\ref{figure:fagnES0msmh} and
\ref{figure:fradioES0msmh}).

One possible solution might be that some S0s have retained a small amount of
tightly bound gas in their cores, which can then be ionized by accretion onto
the black hole.  The absence in radio-bright elliptical galaxies of
H$\alpha$-emitting ionized gas suggests this gas is truly absent -- either
because it is too hot or because it has been expelled.  Gas is heated to X-ray
emitting temperatures during simulations of major mergers \citep[e.g.,][]{Cox06}
-- but the inclusion of hot gas in suites of merger simulations is still
preliminary \citep[see, e.g.,][]{Moster11}, and the results are highly sensitive
to the balance of heating and cooling.

\section{Conclusions}\label{sec:concl}

We have created a catalog of 1064 nearby (median $z \sim 0.02$), bright
($B < 16$ and $\MB < -19$) galaxies, which combines RC3 morphological
classifications and the NYU-VAGC version of the SDSS DR4 photometric and
spectroscopic data; magnitudes have been corrected to account for the
undersubtracted background, and stellar masses were determined using $g - i$
color-based mass-to-light ratios calibrated by \citet{Zibetti09} and applied to
the $i$-band absolute magnitudes. The main morphological classifications
(elliptical vs.\ S0 vs.\ spiral) were checked by visual examination of SDSS
images; a total of 165 galaxies ended up being re-classified, and 55 more were
classified for the first time.

To this dataset we added halo masses and \textit{central} vs.\
\textit{satellite} status (where ``central'' = most massive galaxy in its group)
from the group catalog of \citet[][Y07]{Yang07}, which resulted in a total of
911 galaxies with halo mass assignments; 729 of these are central galaxies and
182 are satellites.  The main advantage of the Y07 catalog is that it spans the
full range of halo masses down to $\Mhalo = 10^{11.63} \, h^{-1} \Msol$. This
allows us to describe the dependence of morphological fractions on halo mass at
$z \sim 0$ for the first time, from smaller clusters down to single-galaxy
halos. (We also determined total number of galaxies per group and group velocity
dispersions for the Y07 groups in our sample, in order to check the robustness
of the halo-mass-based results.) Using the full SDSS DR4 catalog as a parent
sample, we characterized the selection function of our catalog as a function of
$B$ magnitude (synthesized using SDSS colors) and galaxy size ($r_{90}$, the
radius containing 90\% of the Petrosian flux in r-band). Our sample is robust to
a luminosity limit of $\MB = -19.0$, which corresponds to a stellar mass of
$\Mstellar \sim 10^{10.5}\Msol$ for E and S0 galaxies. Galaxies are weighted to
correct for the selection bias, and are also weighted by $V/V_{\rm max}$
to correct for Malmquist bias. This allows us to examine the fraction of
galaxies of various types as a function of stellar and halo mass. We use a
weighted logistic regression method to allow us to assess the statistical
significance of apparent trends. This method has the advantages of modeling a
binomial property without binning of data, and it allows for individual data
points to have weights.

We find that the global fraction of elliptical galaxies increases with
galaxy luminosity and with stellar mass, but not with halo mass. The fraction of
S0s declines to high stellar mass, but increases with halo mass, group velocity
dispersion and number of neighbours. These results are consistent with previous
work at higher redshift which found the S0 fraction to increase in groups
relative to the field population, whilst the elliptical fraction remains roughly
constant \citep{Wilman09}.

The fraction of \textit{central} galaxies with elliptical morphology increases
with stellar and halo mass, consistent with their formation in mergers. In
contrast, the fraction of satellite ellipticals is globally low at all halo
masses, but increases with stellar mass. We interpret this as evidence that
satellite ellipticals were formed as the central galaxies of progenitor halos,
which were subsequently accreted onto their present halo.

Limited to $\Mstellar > 10^{10.5}\Msol$, a modest
fraction of central galaxies are S0s ($20.1 \pm 1.6$\%, with little dependence
on stellar or halo mass). The remaining S0s are satellites of massive halos only
-- we find that the fraction of satellites with S0 morphology rises from
$0^{+2.6}_{-0}$\% in halos with $\Mhalo < 10^{13} \, h^{-1} \Msol$ to
$69.0^{+4.3}_{-4.6}$\% above this threshold.  Presuming S0s to be spirals
in which star formation has been suppressed (leading in turn to the suppression
of spiral arms), we interpret our result as a strong indication that there are
\textit{two} populations of S0s, in which star formation has been suppressed in
different ways.

Central S0s may be formed via suppression of star formation during minor mergers
and/or by feedback from AGN, with similarities to the elliptical population.
Satellite S0s which became S0s while they were still central galaxies within
their progenitor halos constitute a \textit{pre-processed} population which can
account for up to $\sim 20\%$ of the satellite S0s, including all of those with
$\Mstellar \geq 10^{11}\Msol$.

However, the higher fraction of S0s in higher-mass halos implies that many
satellite S0s were accreted as spiral galaxies. These accreted spirals were then
{\it post-processed}, becoming satellite S0s, and are the dominant source of
satellite S0s in the range $10^{10.5}\Msol \leq \Mstellar \leq 10^{11}\Msol$.
Altogether, we estimate that $64 \pm 11$\% of our satellite S0s were accreted as
spirals.

Central S0 and elliptical galaxies frequently host radio sources, consistent
with radio-mode heating of the surrounding hot gas. This heating may offset
cooling onto these galaxies, and thus suppress star formation. However, S0s host
ionized gas components -- mostly LINERs -- much more frequently than elliptical
galaxies of the same mass.

We find no strong dependence of structural subcomponents -- bars, inner
rings/spirals, outer rings -- on environment, in contrast to some earlier
studies \citep[e.g.][]{Elmegreen92}, though we do find evidence that the
frequency of both bars and inner rings increases, and the frequency of outer
rings decreases, with galaxy mass. 

By comparing our results with the study of \citet{Wilman09}, we find
tentative evidence that the fraction of bright S0s in intermediate-mass groups
(those with velocity dispersions $175 \kms \lesssim \sigma \lesssim 500 \kms$)
has increased in the last $\sim 4$ Gyr, rising from $28\pm4$\% at $z \sim 0.4$
to $52.3^{+3.9}_{-4.0}$\%. This is at least qualitatively consistent with
increases in the fraction of S0s in clusters reported by other studies. 

\acknowledgements

We would particularly like to thank our student interns, Mareike Berger and
Daniel Gunzl, for helping check the existing morphological classifications of
the galaxies, and Niv Drory for helping set up the initial database system. We
thank the referee for some very helpful suggestions and comments which
contributed significantly to the improvement of this paper.  We also benefitted
from helpful conversations with, and comments from, Gabriella de Lucia, Fabio
Fontanot, John Mulchaey, Stefano Zibetti, Roberto Saglia, Preethi Nair, Simone
Weinmann, Andrea Biviano, and Dimitri Gadotti.

P.E. was supported by DFG Priority Program 1177 (``Witnesses of Cosmic
History:  Formation and evolution of black holes, galaxies and their
environment'').

Funding for the creation and distribution of the SDSS Archive has been
provided by the Alfred P. Sloan Foundation, the Participating
Institutions, the National Aeronautics and Space Administration, the
National Science Foundation, the U.S. Department of Energy, the Japanese
Monbukagakusho, and the Max Planck Society.  The SDSS Web site is
http://www.sdss.org/.

The SDSS is managed by the Astrophysical Research Consortium (ARC) for
the Participating Institutions.  The Participating Institutions are The
University of Chicago, Fermilab, the Institute for Advanced Study, the
Japan Participation Group, The Johns Hopkins University, the Korean
Scientist Group, Los Alamos National Laboratory, the
Max-Planck-Institute for Astronomy (MPIA), the Max-Planck-Institute for
Astrophysics (MPA), New Mexico State University, University of
Pittsburgh, University of Portsmouth, Princeton University, the United
States Naval Observatory, and the University of Washington.

This research also made use of the Lyon-Meudon Extragalactic Database
(LEDA; http: //leda.univ-lyon1.fr) and the NASA/IPAC Extragalactic
Database (NED); the latter is operated by the Jet Propulsion Laboratory,
California Institute of Technology, under contract with the National
Aeronautics and Space Administration.


\end{document}